\pdfoutput=1
\documentclass[11pt,twoside,a4paper,cmspaper,final,collab]{cms-tdr}
\def\svnVersion{facee12}\def\svnDate{2023/10/20}\def\cmsCernNoTag{CERN-EP-2022-153}\def\cmsCernDate{\today}\def\cmsMessage{Published in Physics Letters B as \href{http://dx.doi.org/10.1016/j.physletb.2023.137713}{\doi{10.1016/j.physletb.2023.137713}.}}
\begin{document}\cmsNoteHeader{B2G-21-004}

\newcommand{\PU}{\HepParticle{U}{}{}\Xspace}
\newcommand{\VLL}{VLL\Xspace}
\newcommand{\VLLL}{\HepParticle{L}{}{}\Xspace}
\newcommand{\VLLE}{\HepParticle{E}{}{}\Xspace}
\newcommand{\VLLN}{\HepParticle{N}{}{}\Xspace}

\newcommand{\ttX}{\ttbar{X}}
\newcommand{\ttbb}{\ttbar{\PQb{\PAQb}}\Xspace}
\newcommand{\ttjj}{\ttbar{jj}}
\newcommand{\Vjets}{\PV{+jets}}
\newcommand{\multiV}{\PV{\PV{(\PV)}}}
\newcommand{\QCD}{QCD multijet\Xspace}

\newcommand{\deltaR}{\ensuremath{\Delta{R}}\Xspace}
\newcommand{\abseta}{\ensuremath{\abs{\eta}}\Xspace}
\newcommand{\Nmisid}{\ensuremath{N^{\text{misid}}\Xspace}}

\newcommand{\RDstar}{\ensuremath{R({\PD}^{*})}\Xspace}
\newcommand{\RD}{\ensuremath{R({\PD})}\Xspace}

\newcommand{\UFO}{universal \textsc{FeynRules} output\Xspace}

\newcommand{\DNNQCD}{\ensuremath{\mathrm{DNN}_{\mathrm{QCD}}}\Xspace}
\newcommand{\DNNTT}{\ensuremath{\mathrm{DNN}_{\ttbar}}\Xspace}
\newcommand{\Njets}{\ensuremath{{N}_{\text{jets}}}\Xspace}

\newcommand{\POWHEGBOX}{\textsc{powheg box}\xspace}
\newlength\cmsTabSkip\setlength{\cmsTabSkip}{1ex}
\newlength\cmsFigWidth
\ifthenelse{\boolean{cms@external}}{\setlength{\cmsFigWidth}{\columnwidth}}{\setlength{\cmsFigWidth}{0.65\textwidth}}

\cmsNoteHeader{B2G-21-004}
\title{Search for pair-produced vector-like leptons in final states with third-generation leptons and at least three \PQb quark jets in proton-proton collisions at \texorpdfstring{$\sqrt{s}=13\TeV$}{sqrt(s) = 13 TeV}}

\author*[cern]{The CMS Collaboration}

\date{\today}

\abstract{
  The first search is presented for vector-like leptons ({\VLL}s) in the context of the ``4321 model'', an ultraviolet-complete model with the potential to explain existing {\PB} physics measurements that are in tension with standard model predictions. The analyzed data, corresponding to an integrated luminosity of 96.5\fbinv, were recorded in 2017 and 2018 with the CMS detector at the LHC in proton-proton collisions at $\sqrt{s}=13\TeV$. Final states with ${\geq}3$ \PQb-tagged jets and two third-generation leptons ($\PGt\PGt$, $\PGt\PGnGt$, or $\PGnGt\PGnGt$) are considered. Upper limits are derived on the {\VLL} production cross section in the {\VLL} mass range 500--1050\GeV. The maximum likelihood fit prefers the presence of signal at the level of 2.8 standard deviations, for a representative {\VLL} mass point of 600\GeV. As a consequence, the observed upper limits are approximately double the expected limits. 
}

\hypersetup{
pdfauthor={CMS Collaboration},
pdftitle={Search for pair-produced vector-like leptons in final states with third-generation leptons and at least three b jets in proton-proton collisions at sqrt(s) = 13 TeV},
pdfsubject={CMS},
pdfkeywords={CMS, vector-like leptons, VLL}}

\maketitle 

\section{Introduction}
\label{sec:intro}

We present a search for vector-like leptons ({\VLL}s), considering the mass range 500--1050\GeV, in the context of the ``4321 model''~\cite{4321_model,benchmark,threeSite_PS,nonuniversal4321}. 
The search uses LHC proton-proton ($\Pp\Pp$) collision data collected by the CMS experiment in 2017 and 2018 corresponding to a total integrated luminosity of 96.5\fbinv~\cite{CMS:2021xjt,CMS-PAS-LUM-17-004,CMS-PAS-LUM-18-002}.

The 4321 model is an ultraviolet-complete model that extends the standard model (SM) gauge groups to a larger $\text{SU}(4) \times \text{SU}(3)' \times \text{SU}(2)_L \times \text{U}(1)'$ group. It is motivated by recent measurements of {\PQb} hadron decays that are in tension with the SM. This tension particularly concerns \RD and \RDstar measurements~\cite{BaBar:2012obs,BaBar:2013mob,Aaij:2015yra,Belle:2015qfa,Belle:2016dyj,LHCb:2017rln,Belle:2017ilt,Aaij:2017uff,Belle:2019rba} that provide evidence for lepton flavour nonuniversality. 
The results of the search described here are the first search results for {\VLL}s in the context of the 4321 model and are also expected to hold for other models with a Pati--Salam leptoquark explaining the B anomalies~\cite{crivellin1,crivellin2}. Final states are probed that are not explored by existing searches for {\VLL}s in models without leptoquarks~\cite{CMS:2019hsm}.

The 4321 model gives a possible explanation for the aforementioned flavour-nonuniversal results, while simultaneously respecting many other measurements that are in agreement with SM expectations and lepton flavour universality~\cite{PhysRevLett.68.3398,Abbott:1999pk,Aaij:2016qqz,Aad:2020ayz}. 
Here, we search for pair production of the lightest new particles in this model, the vector-like leptons. 
These leptons are referred to as \textit{vector-like} because they are non-chiral, \ie their left- and right-handed components have the same charges.

The {\VLL}s come in electroweak doublets with one charged \VLL, {\VLLE}, and one neutral \VLL, {\VLLN}, whose masses are taken to be equal. A mass splitting between these two particles can arise from their couplings to the Higgs boson, but in the 4321 model the splitting is expected to be absent or extremely suppressed. A sizeable mass splitting between the charged and neutral \VLL would affect the \VLL decay modes; therefore a different search strategy would be required. Also, for the model to explain the {\PB} physics anomalies while remaining consistent with other measurements, the mass of the {\VLL}s cannot be too large. In particular, requiring compatibility with the measured \RD and \RDstar anomalies and with measurements of {\PBzs}-{\PABzs} mixing suggests that the \VLL mass cannot be more than a few \TeV~\cite{benchmark,4321footprints}.

\begin{figure*}[h!]
    \centering
    \includegraphics[width=0.24\textwidth]{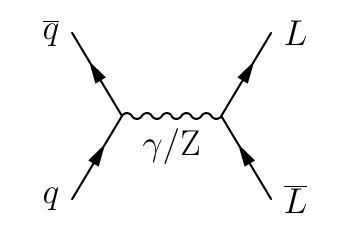}
    \includegraphics[width=0.24\textwidth]{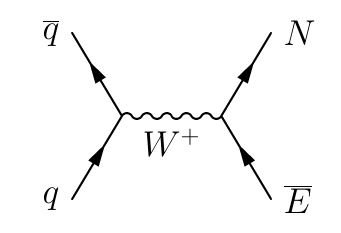}
    \includegraphics[width=0.24\textwidth]{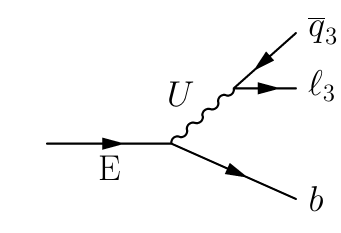}
    \includegraphics[width=0.24\textwidth]{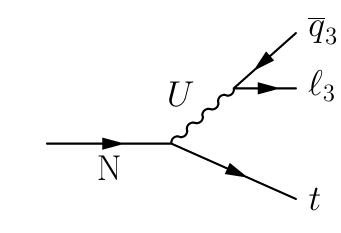}
    \caption{Left side: example Feynman diagrams showing electroweak production of \VLL pairs through $s$-channel bosons, as expected at the LHC. In these diagrams, {\VLLL} represents either the neutral \VLL, {\VLLN}, or the charged \VLL, {\VLLE}. Right side: vector-like lepton decays are mediated by a vector leptoquark, {\PU}. These decays are primarily to third-generation leptons and quarks.}
    \label{fig:feynman_diagrams}
\end{figure*}

The {\VLL}s may be produced via electroweak production through their couplings to the SM {\PW} and {\PZ}/{\Pgg} bosons or via interactions with a new {\PZpr} boson introduced by the 4321 model. In this paper, we consider only electroweak production, and ignore potential contributions from the {\PZpr} boson; the expected size of the {\PZpr} contribution is dependent on the unknown {\PZpr} mass and coupling strength. Despite not explicitly considering this contribution, the results are also expected to hold when including {\PZpr} production. Examples of Feynman diagrams showing the electroweak pair production of {\VLL}s, as well as diagrams of the {\VLL} decays, are shown in Fig.~\ref{fig:feynman_diagrams}. 

The {\VLL}s decay, via a virtual leptoquark, \PU, to two quarks and one lepton. In order to explain the \PB anomalies, the leptoquark is expected to couple most strongly to the third generation and therefore its decays are expected to be almost entirely to third-generation fermions. For each second-generation fermion, a one to two order of magnitude suppression in the branching fraction is expected, and even larger suppressions are expected for any first-generation fermion. For this analysis, we have considered only the dominant decays, \ie decays to third-generation fermions.

The analysis selection is driven by the flavour content of the \VLL decay products. Given the expectation of two third-generation quarks in every \VLL decay, we search for pairs of {\VLL}s by selecting events with a high {\PQb} jet multiplicity. These events are further categorized by the number of {\PGt} leptons. For each {\PGt} multiplicity, dedicated selections are made to divide the category into a signal-enriched ``signal region'' (SR) and one or multiple background-enriched ``control regions'' (CRs). Table~\ref{tab:tau_multiplicities} shows the \PGt multiplicity categories and the decay modes of the different \VLL pairs that contribute to each category. While topologies with electrons or muons in the final state (coming from top quark or tau decays) are possible, we focus only on all-hadronic final states in this analysis. 

\begin{table}[h!]
    \centering
    \caption{ Illustrative contributions from different \VLL production and decay modes to the 0-\PGt, 1-\PGt, and 2-\PGt signal regions. The decay products in parentheses represent the objects coming from the intermediate vector leptoquark, \PU, in the decay. In the table, no distinction is made between particles and antiparticles, the multiplicities of each decay mode are not shown, and the impacts of object misidentification are not considered. The charged and neutral {\VLL}s are represented by {\VLLE} and {\VLLN}; $\mathrm{j}$ represents any quark other than \PQt or \PQb. }
    \begin{tabular}{ccc}
        Tau  & \VLL production & Final \\
		multiplicity & + decay mode & state\\
        \hline 
         \multirow{3}{3em}{0 $\PGt$}& $\VLLE\VLLE \to \PQb(\PQt\PGnGt)\PQb(\PQt\PGnGt)$ & $4\PQb + 4\mathrm{j} + 2\PGnGt$   \\
         & $\VLLE\VLLN \to \PQb(\PQt\PGnGt)\PQt(\PQt\PGnGt)$ & $4\PQb + 6\mathrm{j} + 2\PGnGt$ \\
         & $\VLLN\VLLN \to \PQt(\PQt\PGnGt)\PQt(\PQt\PGnGt)$ & $4\PQb + 8\mathrm{j} + 2\PGnGt$ \\[\cmsTabSkip]
         
         \multirow{4}{3em}{1 $\PGt$} & $\VLLE\VLLE \to \PQb(\PQb\PGt)\PQb(\PQt\PGnGt)$ & $4\PQb + 2\mathrm{j} + \PGt + \PGnGt$ \\
         & $\VLLE\VLLN \to \PQb(\PQt\PGnGt)\PQt(\PQb\PGt)$ & $4\PQb + 4\mathrm{j} + \PGt + \PGnGt$ \\
         & $\VLLE\VLLN \to \PQb(\PQb\PGt)\PQt(\PQt\PGnGt)$ & $4\PQb + 4\mathrm{j} + \PGt + \PGnGt$ \\
         & $\VLLN\VLLN \to \PQt(\PQb\PGt)\PQt(\PQt\PGnGt)$ & $4\PQb + 6\mathrm{j} + \PGt + \PGnGt$\\[\cmsTabSkip]
         
         \multirow{3}{3em}{2 $\PGt$} & $\VLLE\VLLE \to \PQb(\PQb\PGt)\PQb(\PQb\PGt)$ & $4\PQb + 2\PGt$ \\
         & $\VLLE\VLLN \to \PQb(\PQb\PGt)\PQt(\PQb\PGt)$ & $4\PQb + 2\mathrm{j} + 2\PGt$ \\
         & $\VLLN\VLLN \to \PQt(\PQb\PGt)\PQt(\PQb\PGt)$ & $4\PQb + 4\mathrm{j} + 2\PGt$ \\
    \end{tabular}
    
    \label{tab:tau_multiplicities}
\end{table}

A maximum likelihood fit is performed simultaneously across all \PGt multiplicity categories, including both SRs and CRs, to determine the signal strength for each \VLL mass hypothesis. Within each region, differential distributions are used as input to the fit. The distributions are chosen to provide additional separation between the signal and backgrounds. In the fit, the data are also separated by the year in which it was collected (either 2017 or 2018), to ensure a good description of the data-taking and detector conditions in each case.

Machine learning techniques are used to build two deep neural network (DNN) classifiers~\cite{Hastie-Tibshirani-Friedman,Mikuni:2020wpr,Kasieczka:2020pil} to separate signal events from background events. One of the classifiers, \DNNQCD, is trained to distinguish signal from quantum chromodynamic (QCD) multijet events and is used to define the signal region in the 0-{\tauh} category. The other DNN, \DNNTT, is trained to discriminate between signal and \ttbar events and its output distribution is used in the fit for the 1-{\tauh} and 2-{\tauh} categories.

Tabulated results for this analysis are provided in HEPData~\cite{HEPdata}.

\section{The CMS detector and object reconstruction\label{sec:cms}}

The central feature of the CMS apparatus is a superconducting solenoid of 6\unit{m} internal diameter, providing a magnetic field of 3.8\unit{T}. Within the solenoid volume are a silicon pixel and strip tracker, a lead tungstate crystal electromagnetic calorimeter (ECAL), and a brass and scintillator hadron calorimeter (HCAL), each composed of a barrel and two endcap sections. Forward calorimeters extend the pseudorapidity ($\eta$) coverage provided by the barrel and endcap detectors. Muons are measured in gas-ionization detectors embedded in the steel flux-return yoke outside the solenoid. A more detailed description of the CMS detector, together with a definition of the coordinate system used and the relevant kinematic variables, can be found in Ref.~\cite{Chatrchyan:2008zzk}. 

Events of interest are selected using a two-tiered trigger system. The first level, composed of custom hardware processors, uses information from the calorimeters and muon detectors to select events at a rate of around 100\unit{kHz} within a fixed latency of about 4\mus~\cite{Sirunyan:2020zal}. The second level, known as the high-level trigger, consists of a farm of processors running a version of the full event reconstruction software optimized for fast processing, and reduces the event rate to around 1\unit{kHz} before data storage~\cite{Khachatryan:2016bia}. The primary vertex is taken to be the vertex corresponding to the hardest scattering in the event, evaluated using tracking information alone, as described in Section 9.4.1 of Ref.~\cite{CMS-TDR-15-02}.

Reconstruction is performed using the particle-flow algorithm~\cite{CMS-PAPERS-PRF-14-001}, which aims to reconstruct and identify each individual particle in an event with an optimized combination of information from the various elements of the CMS detector. The energy of photons is obtained from the ECAL measurement. The energy of electrons is determined from a combination of the electron momentum at the primary interaction vertex as determined by the tracker, the energy of the corresponding ECAL cluster, and the energy sum of all bremsstrahlung photons spatially compatible with originating from the electron track. The energy of muons is obtained from the curvature of the corresponding track. The energy of charged hadrons is determined from a combination of their momentum measured in the tracker and the matching ECAL and HCAL energy deposits, corrected for the response function of the calorimeters to hadronic showers. Finally, the energy of neutral hadrons is obtained from the corresponding corrected ECAL and HCAL energies.

For each event, hadronic jets are clustered from the reconstructed particles using the infrared and collinear safe anti-\kt algorithm~\cite{Cacciari:2008gp, Cacciari:2011ma} with a distance parameter of 0.4. Jet momentum is determined as the vectorial sum of all particle momenta in the jet, and is found from simulation to be, on average, within 5 to 10\% of the true momentum over the whole transverse momentum (\pt) spectrum and detector acceptance. Additional $\Pp\Pp$ interactions within the same or nearby bunch crossings can contribute additional tracks and calorimetric energy depositions, increasing the apparent jet momentum. To mitigate this effect, tracks identified to be originating from pileup vertices are discarded and an offset correction is applied to correct for remaining contributions~\cite{CMS:2020ebo}. Jet energy corrections are derived from simulation studies so that the average measured energy of jets becomes identical to that of particle level jets. In situ measurements of the momentum balance in dijet, photon+jet, Z+jet, and multijet events are used to determine any residual differences between the jet energy scale in data and in simulation, and appropriate corrections are made~\cite{Khachatryan:2016kdb}. Additional selection criteria are applied to each jet to remove jets potentially dominated by instrumental effects or reconstruction failures. All jets considered in the analysis are required to have $\pt > 40\GeV$ and $\abseta < 2.5$.

To identify jets originating from \PQb quarks, \PQb-tagging algorithms are applied to all reconstructed jets in an event. To identify \PQb jets at the level of the trigger selections, the \textsc{CSV} v2 algorithm is used in the 2017 data sample and the \textsc{DeepCSV} algorithm is used in the 2018 sample~\cite{CMS-PAPERS-BTV-16-002}. In the offline selection and analysis, the \textsc{DeepJet} algorithm is used to identify \PQb jets~\cite{Bols_2020,CMS-DP-2018-058} for both data-taking years.

Each of the tagging algorithms outputs a continuous discriminator value. By requiring this value to be above a given threshold (``working point''), a jet can be identified as a {\PQb}-tagged jet.
The tagging efficiency for \PQb jets is 79 (82)\% for a 1\% misidentification probability of light flavour quark and gluon jets in 2017 (2018); this is referred to as the medium working point (WP).
These values, taken from simulation, are further corrected towards the data using dedicated measurements that correct the identification and misidentification probabilities as functions of the kinematical properties of the jets. In this analysis, the \PQb tagging discriminant score is used as input to the analysis-specific machine learning discriminators, and therefore corrections are applied to the full simulated discriminant score distribution, rather than just to the efficiencies of the individual \PQb-tagging WPs.

Hadronically decaying $\PGt$ leptons ($\tauh$) are reconstructed from jets, using the ``hadrons-plus-strips'' algorithm described in Ref.~\cite{CMS-PAPERS-TAU-16-003}, which combines 1--3 tracks with energy deposits in the calorimeters, to identify the \PGt decay modes.  Modes with two reconstructed tracks target cases in which one of the tracks from a $\PGt$ decay with three charged particles in the final state is not reconstructed. 
To take into account the effects of pair conversion and bremsstrahlung in the material in front of the ECAL, neutral pions are reconstructed by combining electron and photon candidates within a defined region in $\eta$--$\phi$ space, where $\phi$ is the azimuthal angle. The dimensions of this region are in the range 0.05--0.15 in $\eta$ and 0.05--0.30 in $\phi$ , depending on the \pt of the electron and photon candidates.
All {\tauh} candidates in the analysis are required to have $\pt > 20\GeV$ and $\abseta < 2.1$. 
 
 To distinguish genuine $\tauh$ decays from jets originating from the hadronization of quarks or gluons, and from electrons or muons, the \textsc{DeepTau} algorithm is used~\cite{TAU-20-001}. 
Information from all particles reconstructed near the $\tauh$ axis is combined with properties of the $\tauh$ candidate and the event.

Several different WPs of the \textsc{DeepTau} algorithm are used: the very loose WP (in the rest of the paper referred to simply as the ``loose WP``) , medium WP, and tight WP. The rate at which jets are misidentified as a {\tauh} is given in Table~\ref{tab:tau_working_points}, as estimated from simulated samples of \PW boson production in association with jets. 

\begin{table}[ht]
    \centering
    \topcaption{Estimated \tauh identification efficiencies and jet misidentification rates for the \textsc{DeepTau} algorithm used in the analysis.}
    \label{tab:tau_working_points}
	\begin{tabular}{lcc}
        Working point & Efficiency & Jet misidentification rate\\
        \hline
         Loose  &  90\%  & 10\% \\
         Medium &  70\% & 0.4\% \\
         Tight  &  60\% & 0.2\% \\  
	\end{tabular}
\end{table}

The missing transverse momentum vector \ptvecmiss is computed as the negative vector sum of the transverse momenta of all the particle-flow candidates in an event, and its magnitude is denoted as \ptmiss~\cite{Sirunyan:2019kia}. The \ptvecmiss is modified to account for corrections to the energy scale of the reconstructed jets in the event. 
 
\section{Data and simulated samples \label{sec:samples}}

The analysis is performed on a sample of $\Pp\Pp$ collision data with $\sqrt{s}=13\TeV$, collected with the CMS detector in 2017--2018 and corresponding to a total integrated luminosity of 96.5\fbinv~\cite{CMS:2021xjt,CMS-PAS-LUM-17-004,CMS-PAS-LUM-18-002}. The 2017 and 2018 data sets benefit from the upgrade of the pixel tracking detector installed in the winter of 2016--2017~\cite{CMS-TDR-011}. The upgraded tracker significantly improves the {\PQb} jet identification performance~\cite{CMS-DP-2018-033}, which is crucial to the analysis and online event selection. The 2016 data set, collected prior to this upgrade, is therefore not considered.

The signal is modelled at leading order (LO) with \MGvATNLO~\cite{Alwall:2014hca} v2.6.5, using the \UFO interface~\cite{Alloul_2014} to implement the 4321 model. The matrix element calculations are interfaced to \PYTHIA 8.2~\cite{Sjostrand:2014zea}, with the description of the underlying event is provided by the CP5 tune~\cite{CMS-PAPERS-GEN-17-001}. The NNPDF 3.1 next-to-next-to-leading order (NNLO) global fit parton distribution function (PDF) set~\cite{Ball_2017} is used.

Signal production is restricted to use only SM electroweak couplings. We studied the impact of contributions from {\PZ'} production and found that they do not alter our conclusions.  
In the simulated samples, the couplings of the leptoquark are set to 0 for all first- and second-generation SM fermions. The mass of the leptoquark is always taken to be 3.5\TeV and the masses of the two {\VLL}s are always taken to be equal to each other. The simulated model parameters, including the leptoquark and {\PZ'} masses, are taken from benchmark scenarios proposed to explain the \PB anomalies~\cite{4321footprints}. However, due to the much lighter mass scale of the {\VLL}s, the results are mostly insensitive to the values of the boson masses.

We generate signal samples for 12 different {\VLL} mass hypotheses, ranging from 500 to 1050\GeV, in 50\GeV increments. The widths of the {\VLL}s are evaluated for each mass. The decay of the {\VLL}s is performed within \MGvATNLO, however, the decays of any top quarks produced in the {\VLL} decays are simulated with \PYTHIA. 

Simulated samples are also used for background estimation. Samples of \ttbar events are simulated at next-to-leading order using \POWHEGBOX~\cite{Nason:2004rx,Frixione:2007vw,Alioli:2010xd} interfaced to the NNPDF3.1 NNLO PDF set~\cite{Ball_2017}. These samples are generated assuming a 
5-flavour scheme, where {\PQb} quarks are considered massless. The top quark mass is set to 172.5\GeV, and the \texttt{hdamp} parameter, which regulates emissions in \POWHEG, is set to 237.8775\GeV. The inclusive \ttbar production cross section is taken to be 831.76\unit{pb}, as calculated at NNLO in QCD, including resummation of next-to-next-to-leading logarithmic soft gluon terms with \textsc{Top++} 2.0~\cite{Beneke_2012,Cacciari_2012,B_rnreuther_2012,Czakon_2012,Czakon_2013,Czakon_2013.2,Czakon:2011xx}.

Because the signal events are expected to produce four {\PQb} quarks, the \ttbb component of the \ttbar simulation plays a particularly important role in the analysis. This component is taken from the \ttbar simulation described above. Events are taken as part of the \ttbb component if there is at least one jet containing a \PQb hadron not coming from a top quark decay.  Otherwise the event is categorized as \ttjj. 
Recent measurements from both the ATLAS and CMS Collaborations indicate that the cross section of the \ttbb component is underpredicted by the simulation~\cite{CMS-PAPERS-TOP-16-010,atlasttbb,CMS-PAPERS-TOP-18-011,CMS-PAPERS-TOP-18-002}. In order to account for this discrepancy, an overall scale factor of 1.2 is applied to the \ttbb component. 

Smaller backgrounds coming from \ttbar production in association with a boson, denoted {\ttX}, from vector boson production with additional associated jets, denoted {\Vjets}, and from multiboson production, denoted {\multiV}, are also considered. The contributions from these backgrounds are taken from simulation. These processes are all simulated using \MGvATNLO for the matrix element calculation, which is interfaced with \PYTHIA{8} for showering and hadronization.

Samples of QCD multijet events are generated at LO using \MGvATNLO, merging two-, three-, and four-jet matrix elements, and interfaced to \PYTHIA{8} for showering and hadronization. These samples are only used for training the {\DNNQCD} classifier; in the analysis, background contributions from QCD multijet processes are estimated using control samples in data.

For all simulated data, additional minimum bias $\Pp\Pp$ collisions are simulated and superimposed on the primary simulated interaction. Simulation of the passage of particles through the detector material is performed by \GEANTfour~\cite{GEANT4}.

\section{Event selection and categorization \label{categorizaion}}

All events considered in the analysis must satisfy a trigger that requires three {\PQb}-tagged jets each with $\pt > 30\GeV$ and places additional requirements on the \pt of the jets in the event. A further selection, after the trigger, is used in the analysis to avoid the region near the trigger thresholds, which may be sensitive to mismodelling. In order to pass this basic selection, an event is required to contain at least three jets passing the medium {\PQb} tagging WP and at least four jets with \pt greater than 80, 65, 50, and 50\GeV. In addition, the scalar sum of the \pt of all jets in the event (\HT) must be greater than 400\GeV, and no isolated electrons ($\pt > 15\GeV, \abseta < 2.5$) or muons ($\pt > 30\GeV, \abseta < 2.5$) must be present.

\begin{figure*}[ht]
    \centering
    \includegraphics[width=0.75\textwidth]{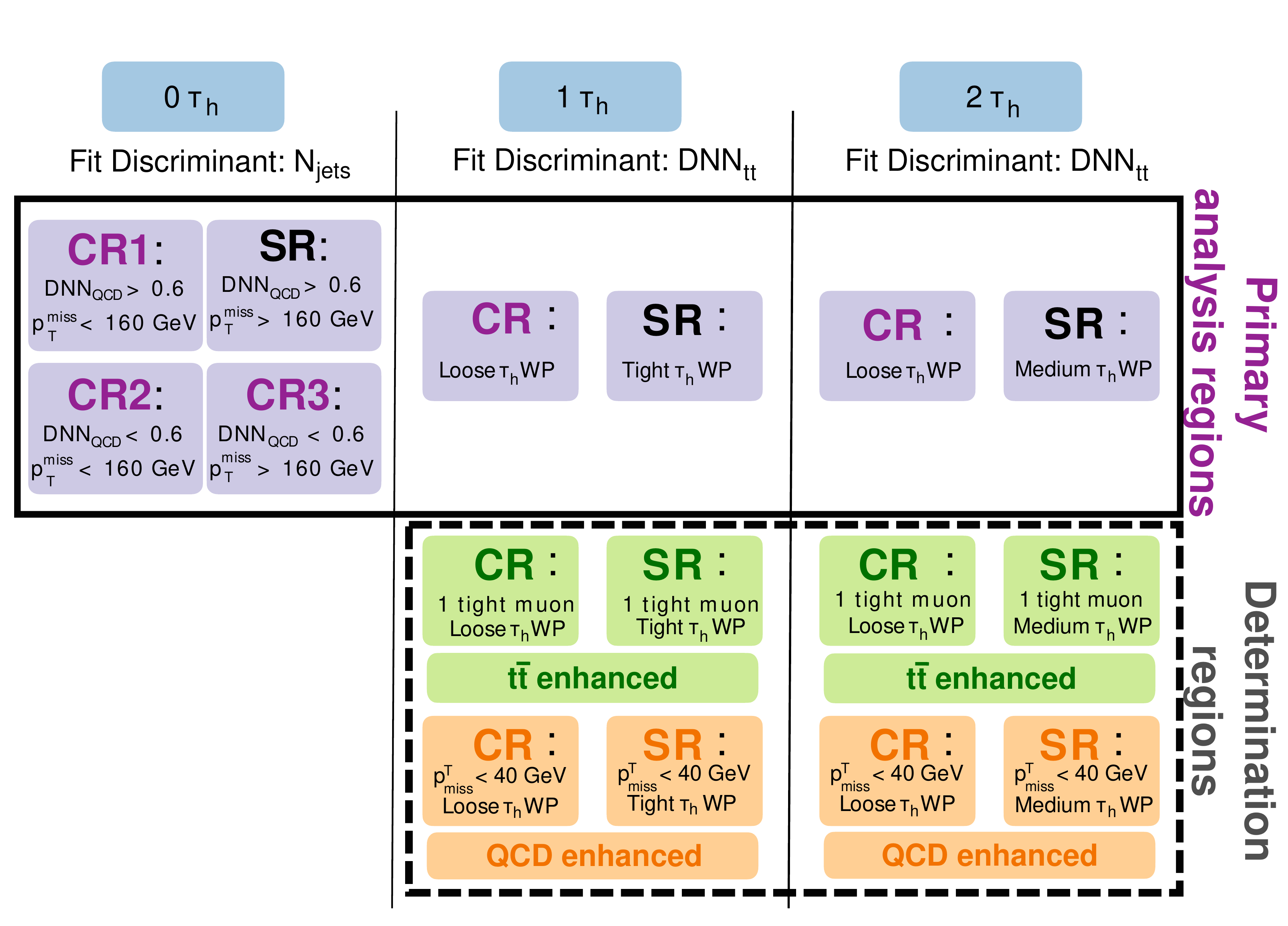}
    \caption{Diagram of the event categorization and the various signal and control regions used in the analysis. The regions within the solid box are all used in the signal extraction fit. The regions in the dashed box are used to determine transfer factors from the control regions to the signal region for events with jets misidentified as {\tauh}s. The selections are mutually exclusive between all regions, so that each event only enters a single region. For brevity, not all selection criteria are shown; the detailed selection criteria are described in the text.}
    \label{fig:regions}
\end{figure*}

These events are then further categorized based on the \tauh candidate multiplicity.  Three categories are considered, with either 0, 1, or $\geq2{\tauh}$ candidates. Within each of these categories, SRs and CRs are defined. The 1-{\tauh} and 2-{\tauh} categories also make use of several determination regions, which are not themselves included in the signal extraction fit, but are used for the background estimation described in Section~\ref{sec:backgrounds}. A diagram summarizing all these regions is shown in Fig.~\ref{fig:regions}.

The 0-{\tauh} region is defined by requiring that there are no {\tauh} candidates passing the loose, medium, or tight WP. No additional event selections beyond the basic ones described above are required. The primary background in this region is from QCD multijet events. This background in the SR is estimated using a technique that makes use of three background-enriched regions in data. The SR is defined by requiring $\ptmiss > 160\GeV$, as expected from the two neutrinos, as well as by making a selection on the {\DNNQCD} (described in Section~\ref{sec:ML}). The jet multiplicity distribution, \Njets, is used for the final fit.

The 1-{\tauh} region is defined by requiring that there is exactly one {\tauh} candidate. 
In the SR, the {\tauh} candidate must pass the tight \tauh WP and there must be no other {\tauh} candidates passing the medium \tauh WP. In the CR, the {\tauh} candidate must pass the loose \tauh WP and there must be no {\tauh} candidates passing the  medium \tauh WP, in order to avoid signal contamination. 
Additionally, in both the SR and the CR, the \tauh candidates must have $\pt > 60\GeV$, and the event must have $\ptmiss>60\GeV$. In the 1-{\tauh} region, significant backgrounds arise from both \ttbar and \QCD events. Multijet events can pass the selection criteria when a jet has been misidentified as a {\tauh}. The \ttbar background has components both with and without genuine {\tauh}s. The \DNNTT classifier distribution, described in Section~\ref{sec:ML}, is used for the final fit.

The 2-{\tauh} region is defined by requiring that there are at least two {\tauh} candidates. In our data set, no events with more than two {\tauh} candidates are found. The SR requires that there are at least two {\tauh} candidates passing the medium WP; the CR requires that there are at least two {\tauh} candidates passing the loose WP and no {\tauh} candidates passing the medium WP, in order to avoid signal contamination. Additionally, for all events in 2-{\tauh} region, the {\tauh} candidates must have opposite charge, and be separated by $\deltaR =  \sqrt{\smash[b]{(\Delta\eta)^2 + (\Delta\phi)^2}}$  of at least 0.5.
The event is required to have $\ptmiss>40\GeV$. In this region, \ttbar is the dominant background and the QCD multijet contribution is highly suppressed. 
The same \DNNTT used in the 1-{\tauh} region is also used in the final fit of the 2-{\tauh} region.

\section{Machine learning classifiers \label{sec:ML}}

Two DNN classifiers are trained to separate signal and background events. One of them, \DNNQCD, is specifically trained to discriminate between signal and \QCD events. The other, \DNNTT, is trained to distinguish signal from \ttbar events. 

Both DNNs use the same underlying architecture, a graph neural network called \textsc{ABCNet}~\cite{Mikuni:2020wpr}. In this implementation, individual final state objects are passed as inputs to the DNN, and are represented as nodes in the graph. The graph connects sets of $\mathrm{k}$-nearest neighbour objects with edges, so that relationships between the objects are taken into account. In all cases, we use $\mathrm{k}=8$ and a Euclidean metric in the $\eta$--$\phi$ coordinates. 

Both {\tauh} candidates and jets are passed to \textsc{ABCNet} as inputs, with up to ten objects per event used as input. If more than ten objects are present in the event, all of the {\tauh} candidates are passed to the network first, followed by the jets, ordered by \pt, until the total object count has reached ten.

For each object, several kinematic features are passed to \textsc{ABCNet}. They are shown in Table~\ref{tab:ML_inputs}.

\begin{table}[ht]
    \centering
    \topcaption{List of input features used during training of the \textsc{ABCNet} model for \VLL classification.}
    \label{tab:ML_inputs}
	\begin{tabular}{ll}
        Variable & Description\\
        \hline
         $\eta$       &  Pseudorapidity\\  
         $\phi$       &  Azimuthal angle \\
         $\log(\frac{\pt}{\GeV})$  &  Logarithm of the \pt \\  
         $\log(\frac{m}{\GeV})$    &  Logarithm of the mass \\  
         $Q$          &  Charge \\  
         \textsc{DeepJet} score      &  \PQb tagging discriminant \\  
	\end{tabular}
\end{table}

For jets, the charge entry is always set to zero. For $\tauh$ candidates, the \textsc{DeepJet} score is always set to zero, which corresponds to an extremely low probability of being a \PQb jet. With this information, the model is able to distinguish $\PGt$ leptons from jets during training. Using the same model, two different trainings are performed to distinguish \VLL events from QCD and \ttbar events. An equal mixture of signal samples with different mass hypotheses is used in the DNN training. The simulated events used for training the {\DNNQCD} classifier are required to pass the 0-{\tauh} selection criteria, whereas those used for training the {\DNNTT} classifier must have at least one {\tauh} candidate passing the medium WP.

\section{Background estimation\label{sec:backgrounds}}

The primary SM processes that pass the event selection and form backgrounds in the search are \ttbar and \QCD production. 
Additional contributions from {\ttX}, {\Vjets}, and {\multiV} production are also considered, but play smaller roles.
The estimated contributions of these minor background processes are taken directly from simulation, and the {\Vjets} and {\multiV} contributions are found to be negligible. On the other hand, the {\QCD} contribution is estimated using control samples in data.

The expected contributions from {\ttbar} events, including the {\ttbb} component, are estimated mostly from simulation, as described in Section~\ref{sec:samples}. 
The number of {\ttbar} events with misidentified {\tauh} candidates, \ie where at least one identified {\tauh} candidate does not come from a true hadronic $\PGt$ lepton decay, is estimated from a combination of simulation and control samples in data. The estimation is performed by means of a ``fake factor'' method. The method uses a CR with a looser {\tauh} WP to estimate the number of events in the SR with a tighter {\tauh} WP. The ratio of the number of events in the two regions, called the fake factor ($\text{FF}$), is first measured in a dedicated determination region ($\text{DR}$) which is orthogonal to the primary analysis region. Then in the primary analysis region, the estimated number of events in the control region is multiplied by the fake factor to estimate the number of events in the SR:
\begin{linenomath}
\begin{equation}
\begin{aligned}
	\text{FF} &\equiv \frac{\Nmisid(\text{SR}_{\mathrm{DR}})}{\Nmisid(\text{CR}_{\mathrm{DR}})}, \\
	\Nmisid(\text{SR}) &=  \Nmisid(\text{CR})\text{FF},
\end{aligned}
\end{equation}
\end{linenomath}
where $\Nmisid(X)$ is the number of events in region $X$ with at least one misidentified {\tauh} candidate. The FF and {\Nmisid} are estimated in bins of the {\DNNTT} score.

To estimate {\Nmisid} in the DRs, the number of events in data with correctly identified {\tauh} objects is estimated from simulation and subtracted from the total; the remaining component is taken to be \Nmisid. In the primary analysis region, $\Nmisid(\text{CR})$ is taken directly from simulation. An additional uncertainty is applied to this component to account for possible mismodelling. 

The DRs are designed to be enriched in \ttbar events. After passing a single-muon trigger, these events are required to pass the same selection as the main analysis, except that they must have at least two (not three) {\PQb}-tagged jets. Events are categorized based on the number of \tauh candidates as in the main analysis region. However, unlike in the main analysis regions, no additional selection criteria (on di-$\PGt$ charge, $\tauh$ \pt, or \ptmiss) are applied when performing the \tauh categorization in the DRs. These differences in selection are all designed to increase the number of events in these regions, as the primary limitation of our FF estimation is the statistical uncertainty. An additional correction and uncertainty, to account for these differences in the phase space between the determination and primary analysis region, is estimated from simulation. 

The fake factor method is applied independently for the 1-{\tauh} and 2-{\tauh} regions, using the same methodology in both cases.

The {\QCD} background in the 1-{\tauh} and 2-{\tauh} regions is also determined using a fake factor method. Unlike the approach adopted for {\ttbar}, where $\Nmisid(\text{CR})$ is taken from simulation, this component is also taken from data for the {\QCD} background. The number of expected events for all other background processes is subtracted from the data to estimate $\Nmisid(\text{CR})$.  The signal contribution is ignored in the subtraction, since it is never expected to contribute more than $\mathcal{O}(1\%)$ of the \QCD yield in the control region. The FF DR is constructed by requiring $\ptmiss<40\GeV$. As for the \ttbar fake factor, a correction is applied to account for the difference between the FFs in the DR and the primary analysis region. However, in the 2-{\tauh} region, no simulated {\QCD} events pass the signal region selection criteria and therefore no correction is applied.

In the 0-{\tauh} region, the {\QCD} contribution is estimated using an ``ABCD'' method. This method makes use of two observables, over which the background distribution is uncorrelated. Each observable is divided into two subregions, to form four regions in total. Then, from the fact that the two observables are uncorrelated, the background contribution in a signal-enriched region is estimated from the background yields in three signal-depleted regions:
\begin{equation}
    N^{\mathrm{QCD}}(\text{SR}) = \frac{N^{\mathrm{QCD}}(\text{CR1})N^{\mathrm{QCD}}(\text{CR3})}{N^{\mathrm{QCD}}(\text{CR2})},
\end{equation}
where $N^{\mathrm{QCD}}(X)$ represents the number of \QCD events in region $X$ and is determined independently for each bin of the {\Njets distribution used in the fit. We use the \ptmiss and the {\DNNQCD} classifier score as the two observables for the ABCD method. The regions are divided at $\DNNQCD = 0.6$ and $\ptmiss = 60\GeV$, with the region $\DNNQCD > 0.6, \ptmiss > 60\GeV$ being the SR. The training of the DNN is designed to ensure that its output and the \ptmiss are uncorrelated~\cite{Kasieczka:2020pil}, and this assumption was further tested both in simulated samples and in a dedicated validation region.

\section{Uncertainties \label{sec:uncertainties}}

The dominant uncertainty comes from the limited number of events in the 1-{\tauh} and 2-{\tauh} SRs. The statistical uncertainties are approximately 70\% of the total uncertainty in the signal strength estimate. A number of other experimental and theoretical uncertainties are considered and included in the fit as nuisance parameters. In each case, the nuisance parameters alter one or more of the processes included in the fit. Except where otherwise noted, the nuisance parameters alter both the shape and normalization of the affected processes. 

Uncertainties in the measurement of the energies and resolutions of jets and {\tauh} candidates are estimated from the calibrations described in Section~\ref{sec:cms}. 

Uncertainties in the {\PQb}-tagging efficiencies and misidentification rates are taken from dedicated measurements~\cite{CMS-PAPERS-BTV-16-002}. The tagging rates and their uncertainties are determined by looking separately at events enriched in genuine {\PQb} jets, using a \ttbar selection, and events enriched in light flavour quark and gluon jets, using $\PZ \to \Pe^+\Pe^-$ and $\PZ \to \PGm^+ \PGm^-$ events with additional jets.  An iterative fit to these two regions is performed and the uncertainties are characterized by nine separate components, which account for different uncertainty sources such as the limited sample size and uncertainties in the flavour compositions of the regions.

Uncertainties in the identification and misidentification rates of {\tauh} candidates are also considered. The identification rate uncertainties are taken from measurements using a set of $\PZ \to \PGt\PGt$ enriched events, where one of the $\PGt$ leptons decays into neutrinos and a muon~\cite{CMS-PAPERS-TAU-16-003}. The muonically decaying $\PGt$ lepton is used as a probe to measure the identification rate for the hadronically decaying $\PGt$ lepton. For each {\tauh} identification WP, there is one nuisance parameter that alters the {\tauh} identification efficiency. 

Three additional sets of uncertainties related to {\tauh} misidentification are included. A 20\% normalization uncertainty is applied to the contributions arising from misidentified {\tauh} candidates in the CRs, \ie where the loose {\tauh} WP is used. To account for the misidentification uncertainties at tighter WPs, the uncertainties from the FF derivation are propagated to background components with misidentified {\tauh} candidates in the signal regions, \ie where the medium or tight {\tauh} WP is used. 
Systematic uncertainties in the simulated samples used in the derivation of the FFs are small compared with the associated statistical uncertainties, and are neglected. 
Each bin in which FFs are derived is therefore assigned an independent uncertainty corresponding to the statistical uncertainty in the FF derivation, which is typically between 10\% and 50\%. 
There are 20 such uncertainties for the {\QCD} FFs and nine for the \ttbar FFs. 
Finally, a set of uncertainties accounts for differences between the FFs in the determination regions and the main fit region. 
The uncertainty is taken to be the relative difference between the two FFs, as found in simulation, and is comparable to the statistical uncertainty in the FFs. For the 2-{\tauh} {\QCD} FFs, where no simulated events pass the selection criteria, an 80\% uncertainty is applied. These uncertainties are fully correlated within each region and across years.
The total FF uncertainties are the second largest source of uncertainty in the measurement and are approximately 40\% of the uncertainty in the signal strength estimate.		

A simplified Barlow--Beeston approach~\cite{Barlow:1993dm}, with a single nuisance parameter per bin of the distribution, is used to treat the statistical uncertainties due to the limited size of the simulated samples.

Other experimental uncertainties that have a smaller impact on the analysis include: uncertainties in the total delivered luminosity (normalization only), the uncertainty in the number of inelastic $\Pp\Pp$ collisions in a single bunch crossing, uncertainties related to detector conditions, and uncertainties in the determination of the trigger efficiencies.

Theoretical uncertainties are assessed separately for the \ttbar, \ttX, and signal processes. Each of these uncertainty sources is treated as uncorrelated between these processes. For each process, a single PDF uncertainty is constructed by adding in quadrature the contributions from all of the Hessian eigenvectors of the process PDF set and its variations in the strong coupling (\alpS) ~\cite{Butterworth:2015oua}. Independent variations of the renormalization and factorization scales by factors of two are used to assess the uncertainty from the approximations used in the matrix element calculations; an envelope that describes the largest impacts of these variations is formed and encoded as a single nuisance parameter per process. Variations where the two scales differ by a factor of four are not included. Uncertainties in the parton shower simulation are assessed by changing the scale at which \alpS is evaluated in the shower by factors of two up and down. This is done independently for the initial state and final state radiation, creating two independent nuisance parameters for each process. An additional uncertainty related to the mismodelling of the top quark \pt spectrum in \ttbar events by next-to-leading order Monte Carlo generators is also applied to the \ttbar sample. 

An additional 20\% normalization uncertainty is applied to the \ttbb component. This accounts for the fact that the \ttbb component is not as well determined as the overall \ttbar cross section.

\section{Fit setup and validation\label{sec:fit}}

The results of the analysis are extracted by performing a simultaneous binned maximum likelihood fit over both years and all signal and control regions. The SRs defined in Fig.~\ref{fig:regions} provide the main sensitivity to any potential signal, while the CRs are used to provide estimates of some of the important background processes from data. For the 0-{\tauh} region, the distribution of the number of jets in the event (\Njets) is used in the fit, whereas in the 1-{\tauh} and 2-{\tauh} regions the distributions of the {\DNNTT} scores are fitted.

The fit is performed separately for each mass hypothesis to extract any dependence of the result on the \VLL mass. In each case, a single signal strength parameter that multiplies the \VLL production cross section is used. Systematic uncertainties are treated as nuisance parameters in the fit. Nuisance parameters that affect the overall rate of a process are described by log-normal probability density functions, whereas nuisance parameters that affect the shapes of distributions are described by Gaussian probability density functions. Upper limits at the 95\% confidence level (\CL) on the signal strength are determined using the \CLs criterion~\cite{CLS2,CLS1} with the profile likelihood ratio as the test statistic. In this procedure, we make use of the asymptotic approximation~\cite{Cowan:2010js}. The signal significance is calculated from the likelihoods of the background-only and signal-plus-background models using Wilks' theorem~\cite{wilks1938}.

The analysis procedure and optimization is performed ``blinded'' (without looking at the data) and a number of validation tests are also performed on the fit setup without using the observed data in the main fit region. Fits are performed on pseudo-data generated based on the expected background model, and then applying Poisson-distributed smearing to the values in each bin. This test is performed in four configurations: without signal, and also including signal simulation with three different input cross sections. For each configuration, 500 pseudo-experiments are performed. The signal strength extracted from the fits is consistent with it being an unbiased estimator of the true signal strength, and the uncertainties quoted on the signal strengths are consistent with the expected frequentist coverage.

Fits are also performed on collision data in validation regions. These regions are chosen to be orthogonal to the primary analysis regions, and with almost no signal contamination. Separate validation regions are chosen for each of the 0-{\tauh}, 1-{\tauh}, and 2-{\tauh} categories. In every case, the validation region is chosen to have exactly two \PQb-tagged jets (as compared to $\geq3$ in the main analysis region). In the 1-{\tauh} and 2-{\tauh} categories, the validation regions also require that the event score from the DNN classifier separating signal from \ttbar is less than 0.8, in order to ensure that there is no signal contamination. The results of these fits are consistent with the background-only hypothesis.

\section{Results \label{sec:results}}

Postfit distributions for the background-only and signal-plus-background fits are shown in Fig.~\ref{fig:postfit}. For the fits including signal, the 600\GeV\ {\VLL} mass point is shown as an example. The signal fit shows consistency across the sensitive channels. 

 The observed data show excesses in the highest {\DNNTT} bins for both the 1-{\tauh} and 2-{\tauh} channels, for both 2017 and 2018. The 0-{\tauh} channel does not show any excess, which is expected given its lower sensitivity.

\begin{figure*}
    \centering
    \includegraphics[width=0.29\textwidth]{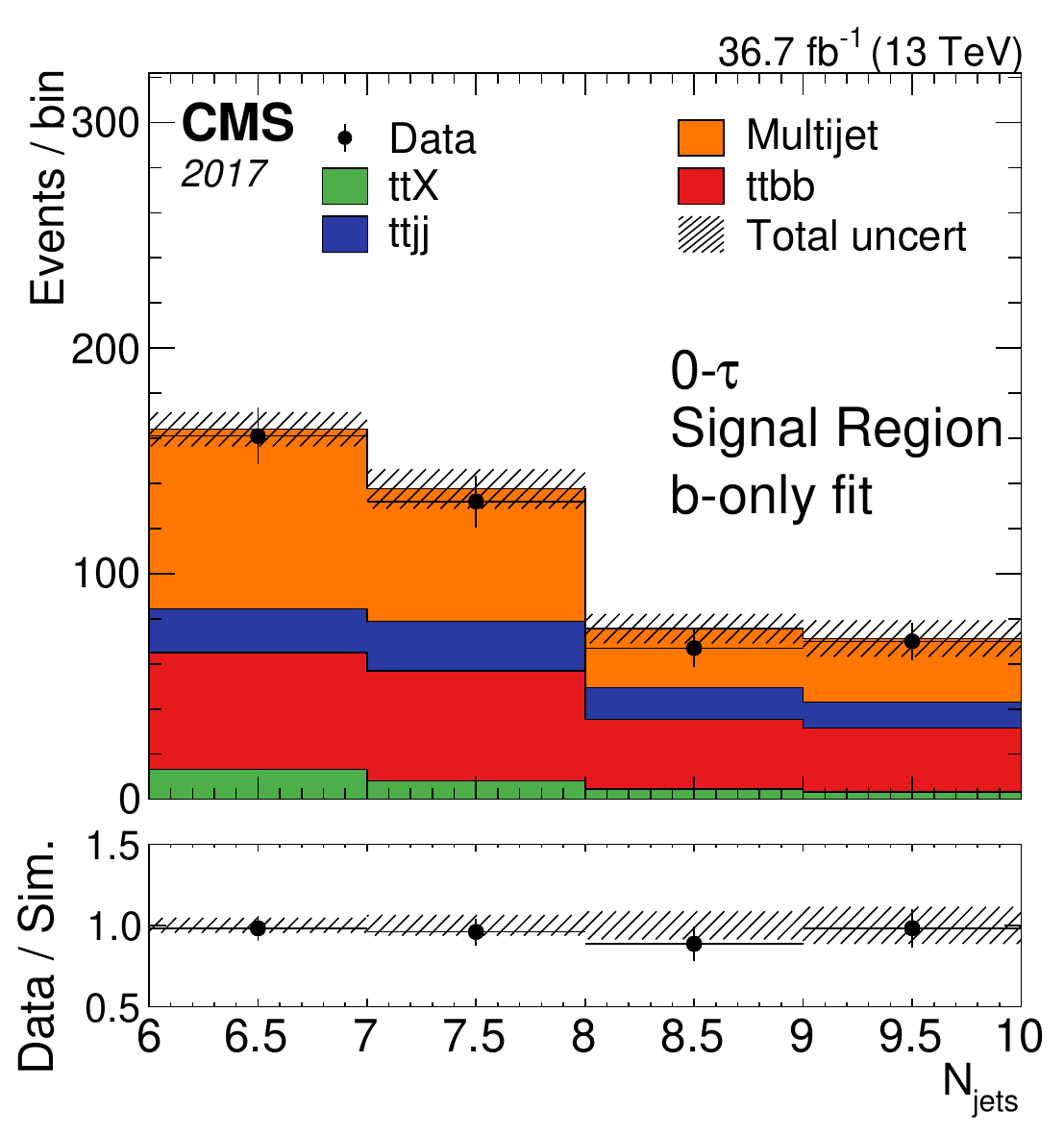}
    \includegraphics[width=0.29\textwidth]{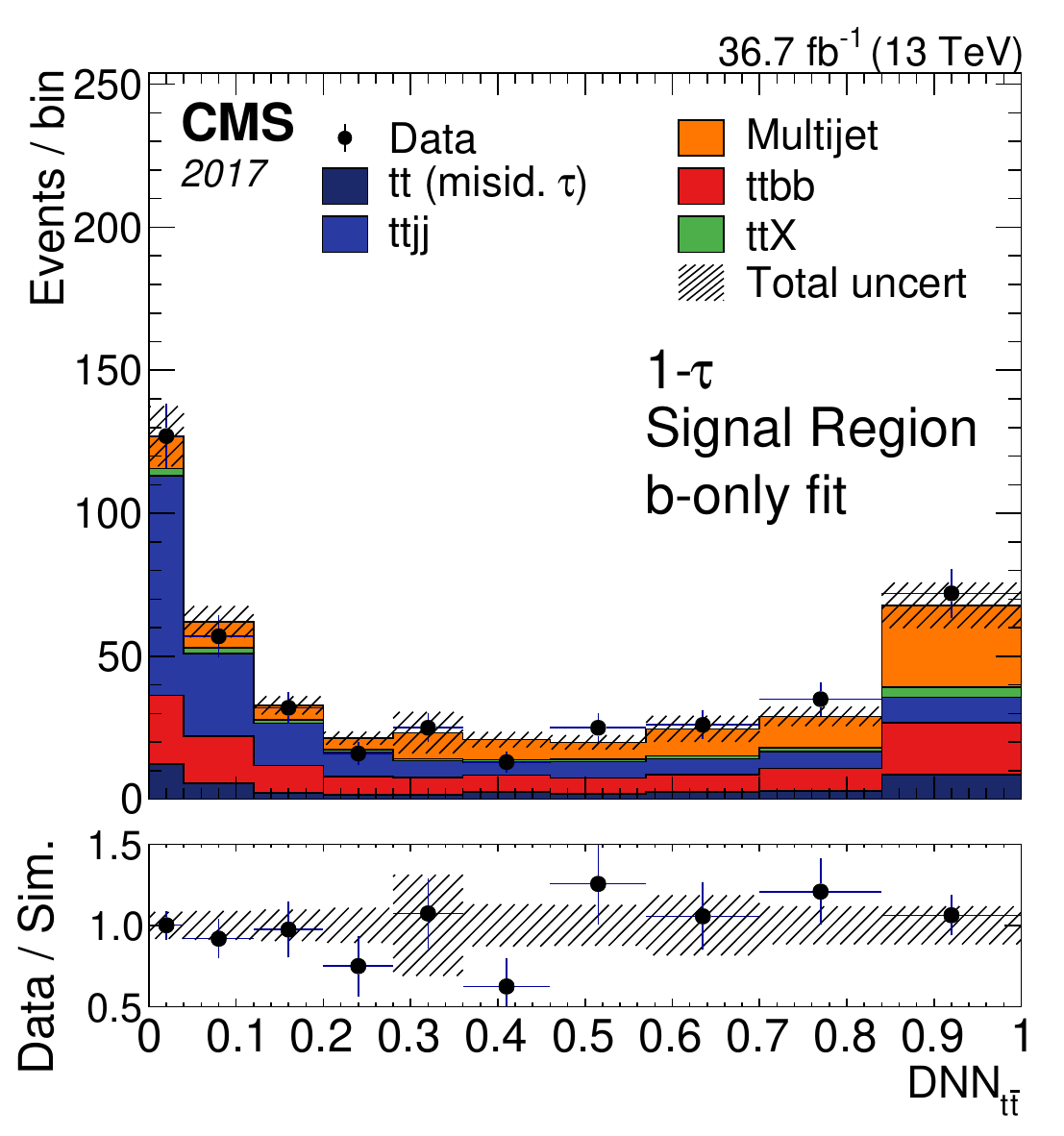}
    \includegraphics[width=0.29\textwidth]{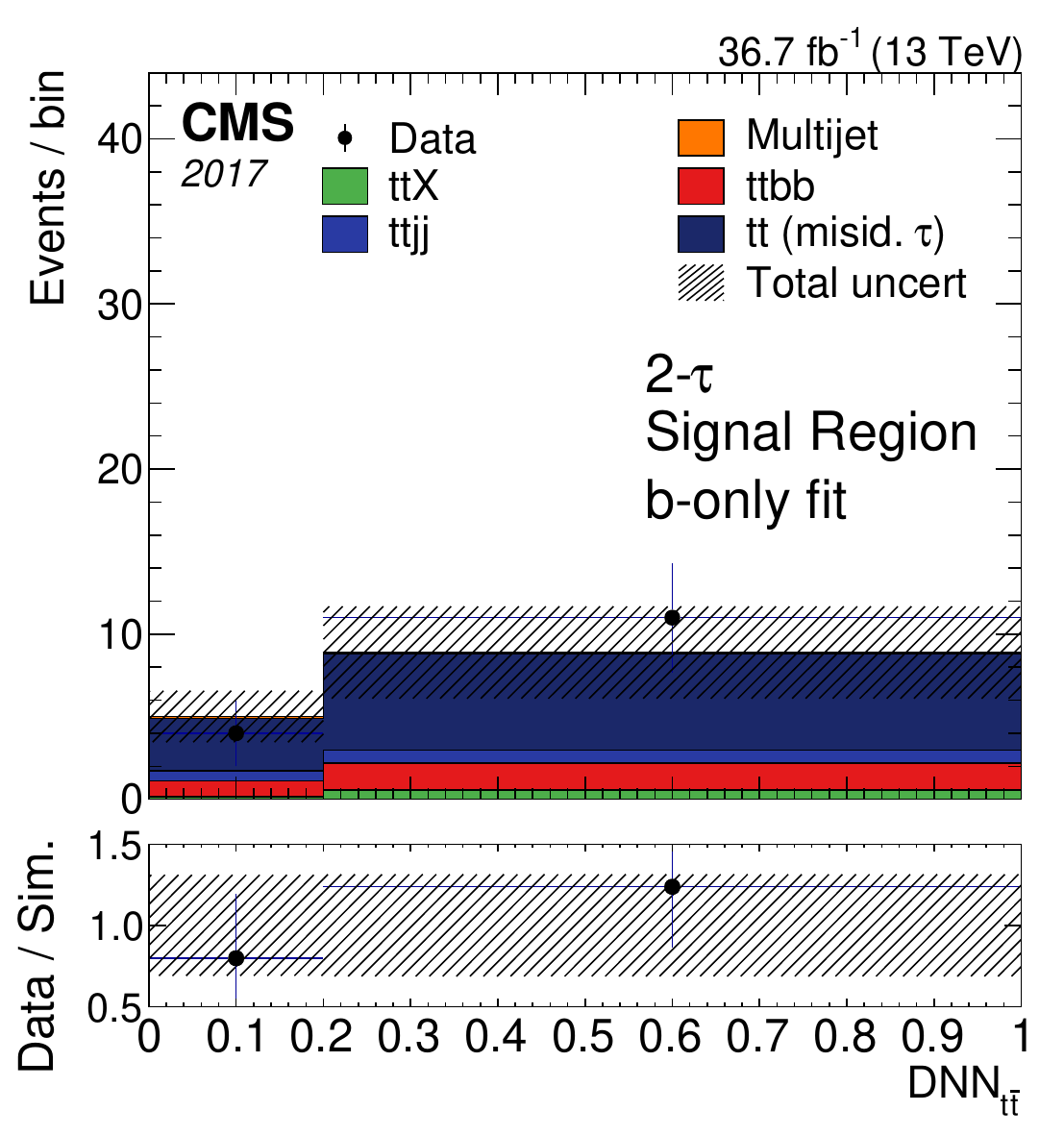}

    \includegraphics[width=0.29\textwidth]{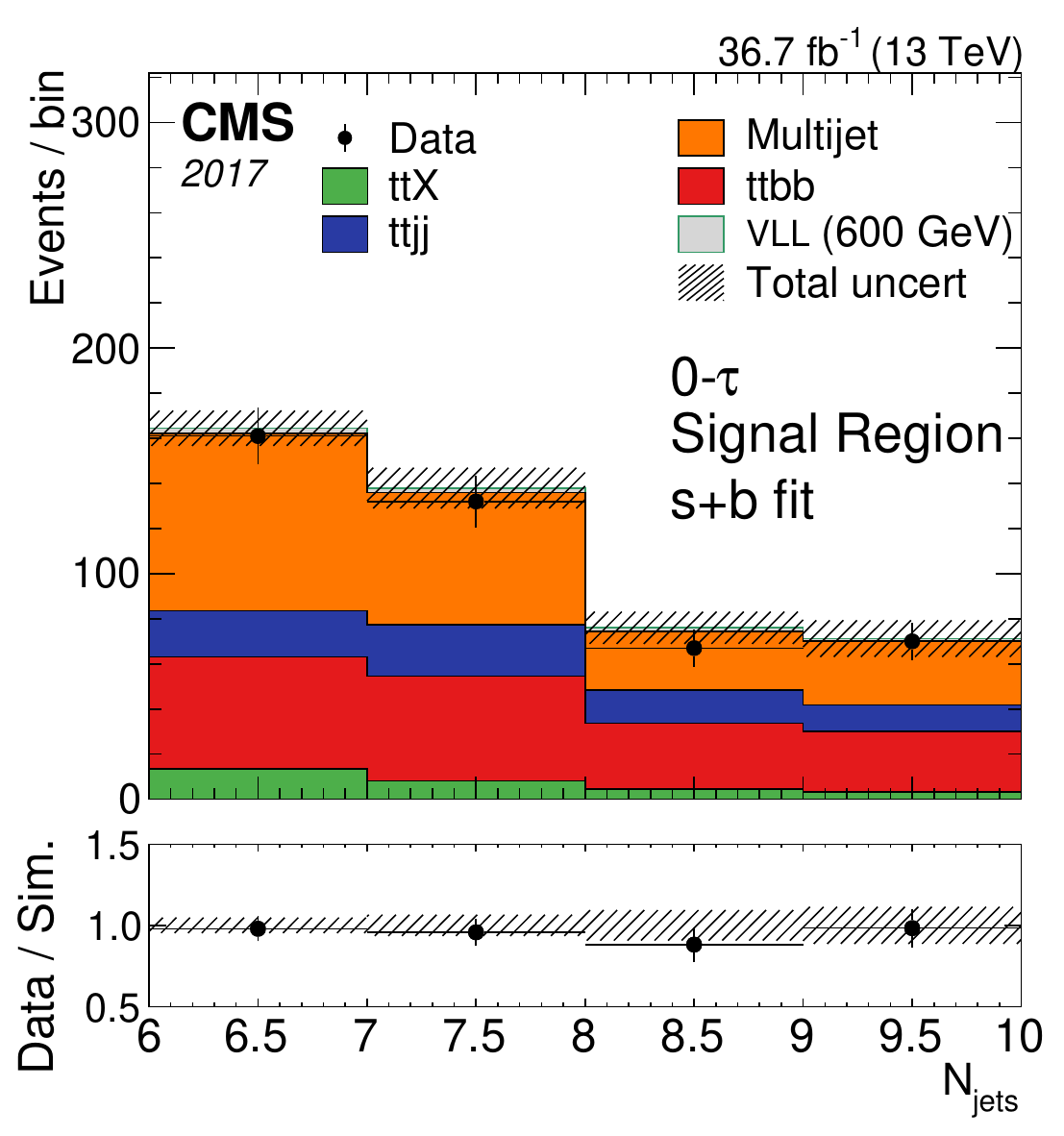}
    \includegraphics[width=0.29\textwidth]{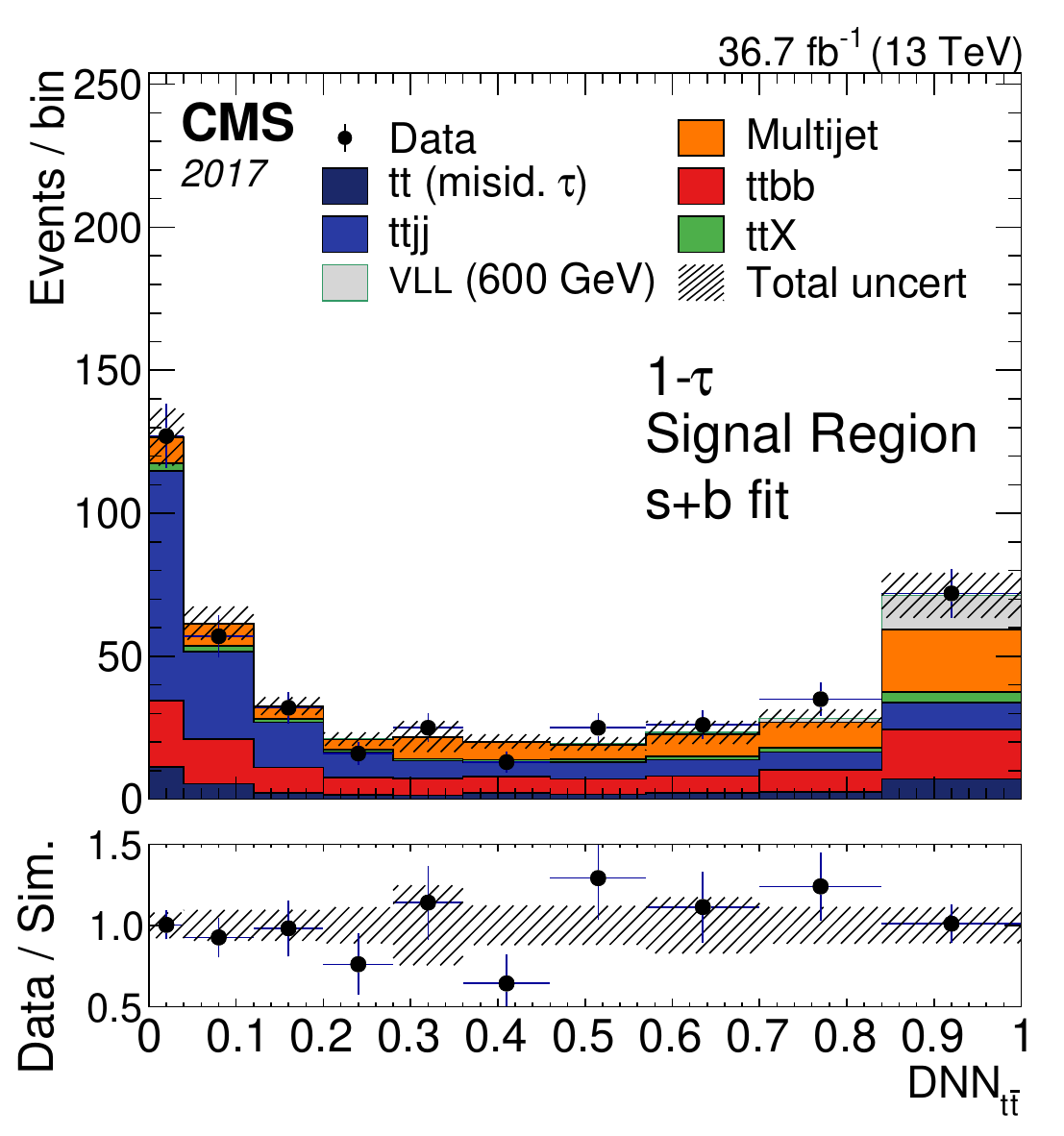}
    \includegraphics[width=0.29\textwidth]{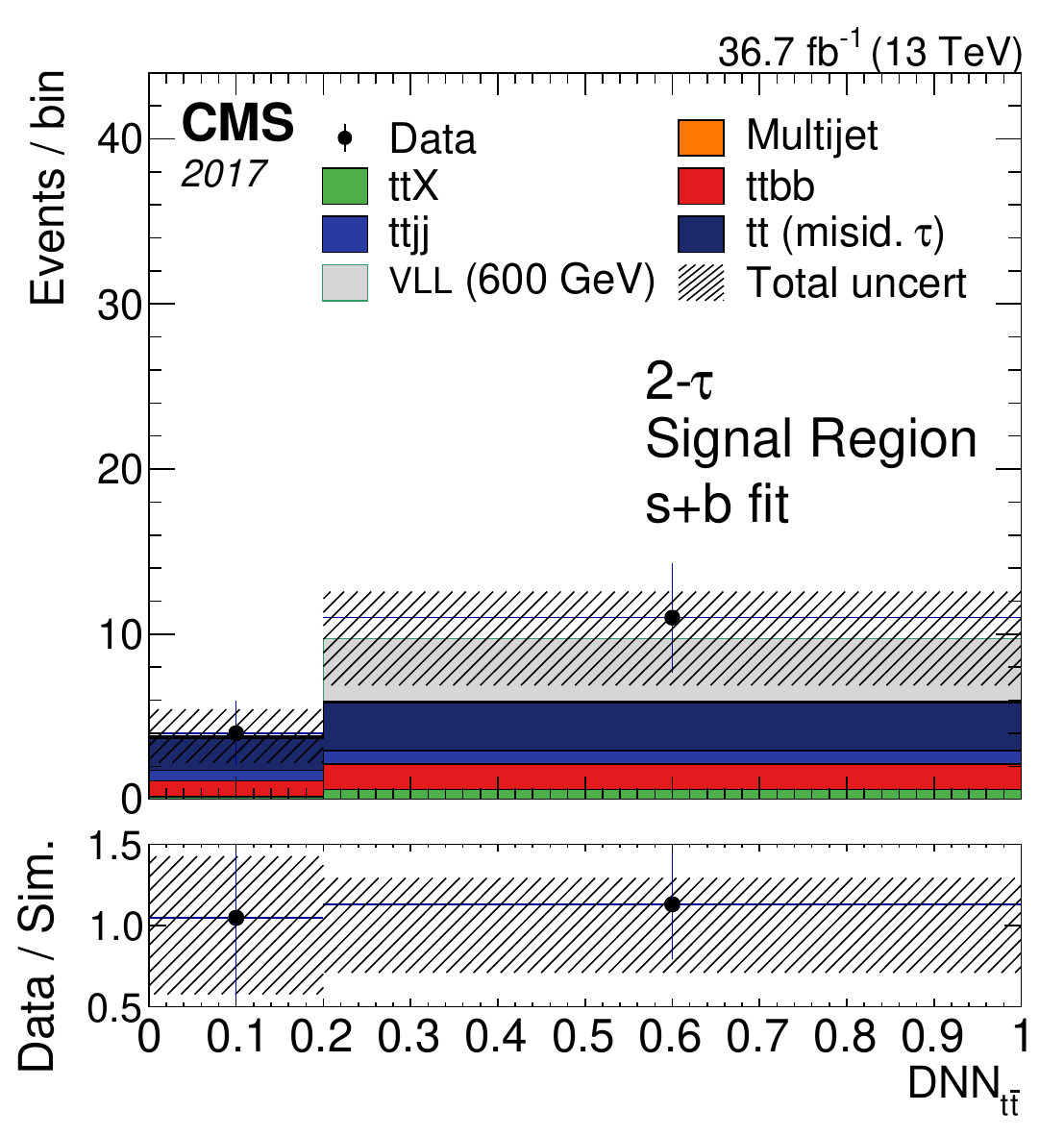}

    \includegraphics[width=0.29\textwidth]{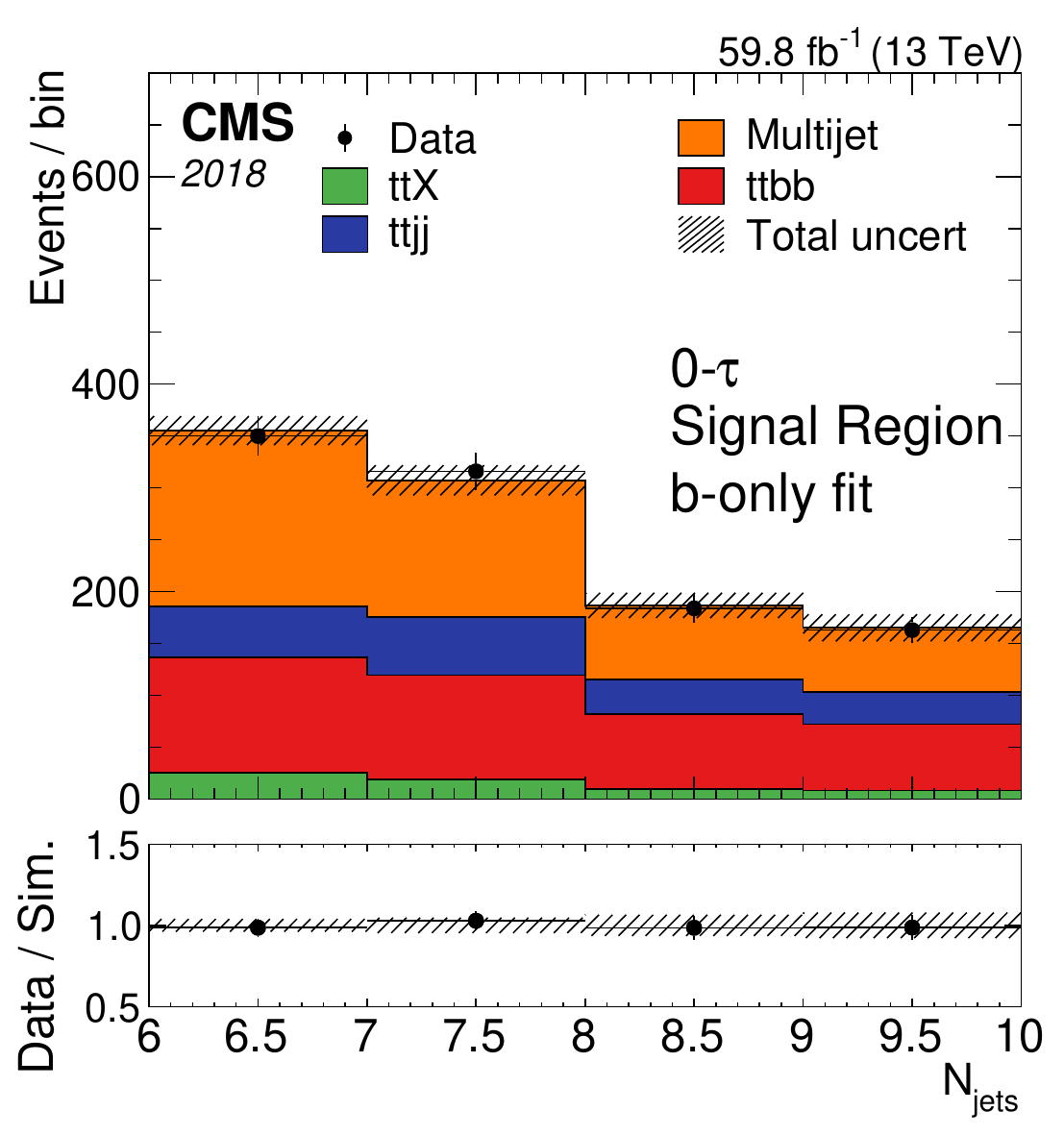}
    \includegraphics[width=0.29\textwidth]{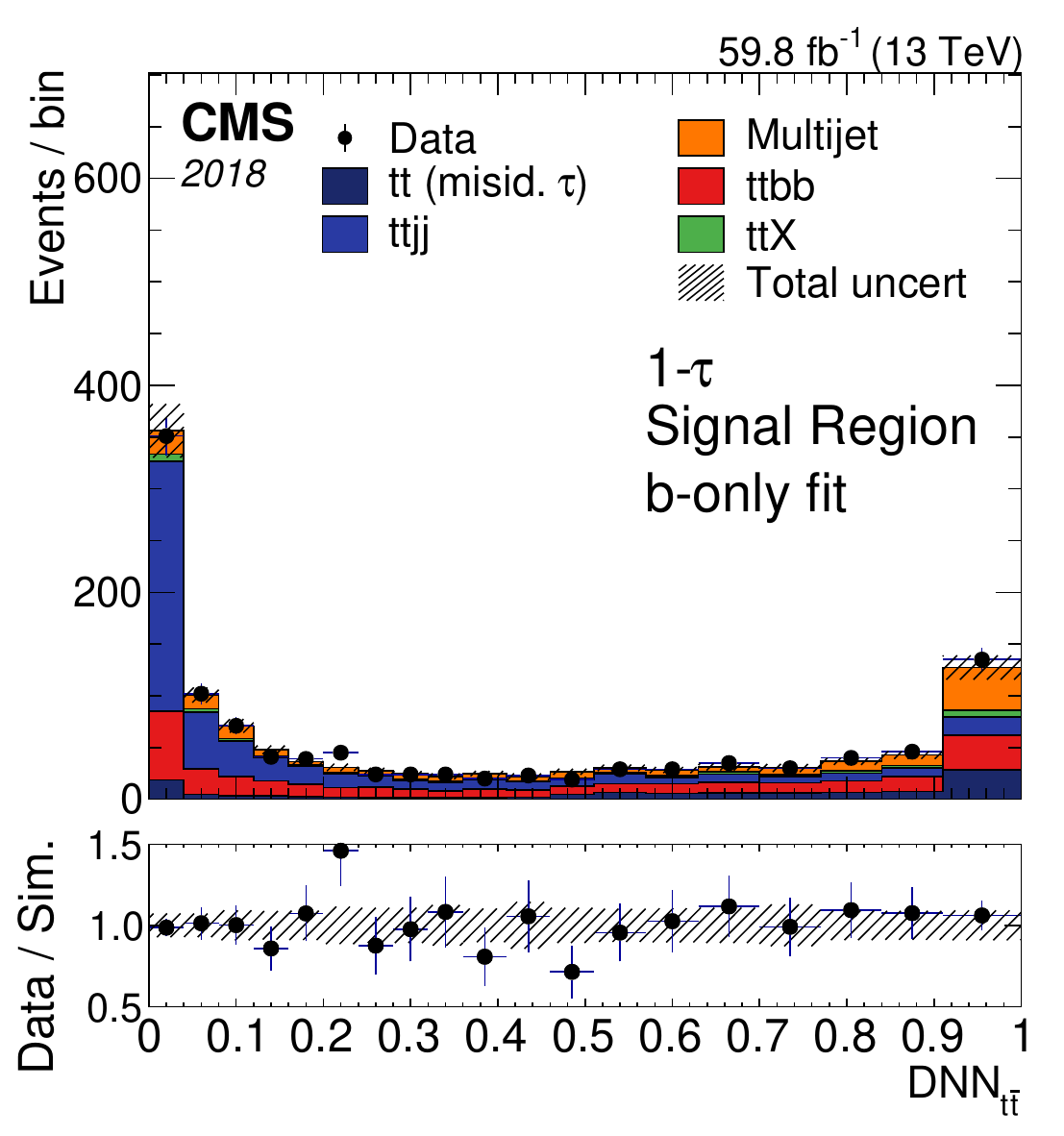}
    \includegraphics[width=0.29\textwidth]{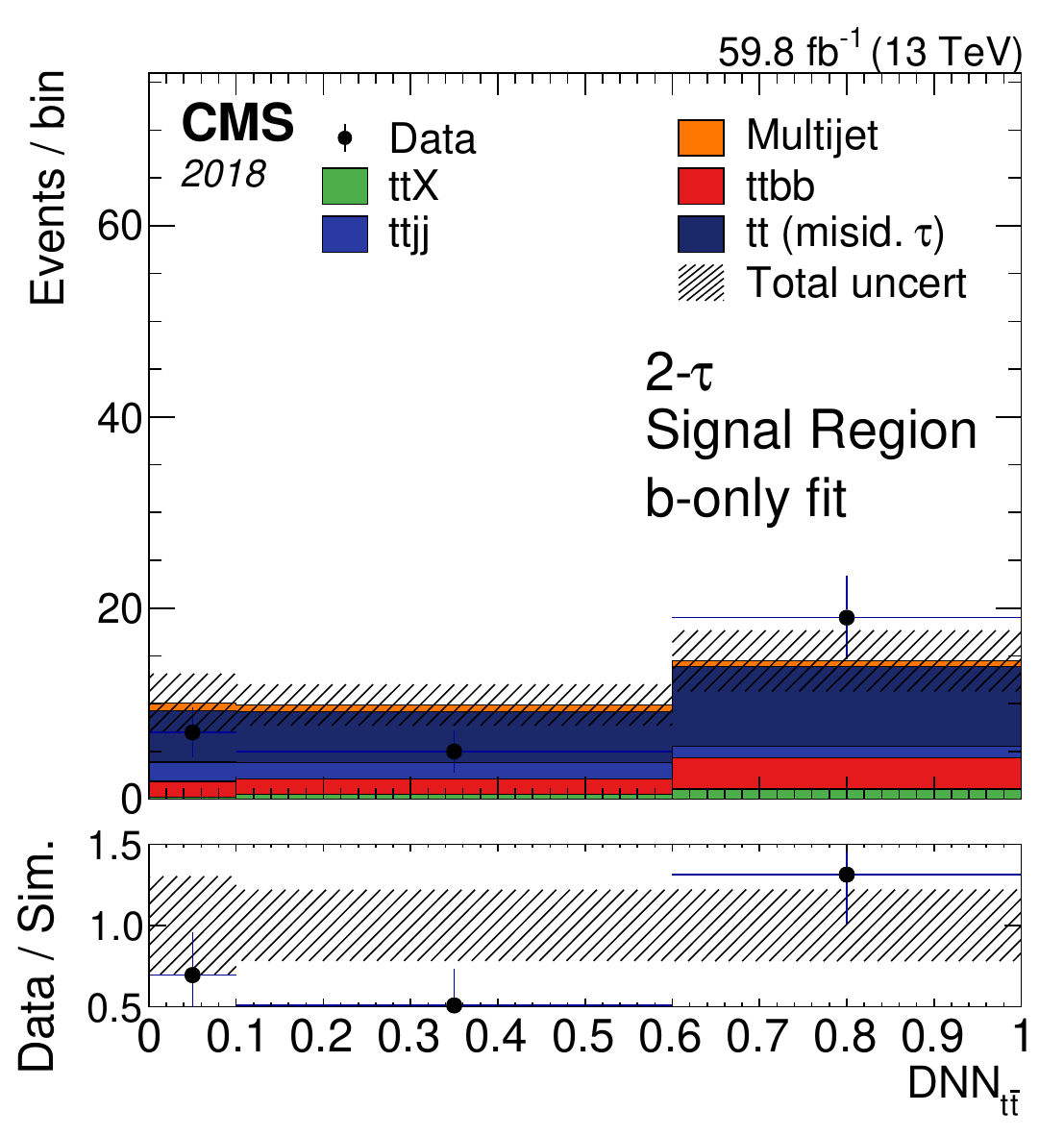}

    \includegraphics[width=0.29\textwidth]{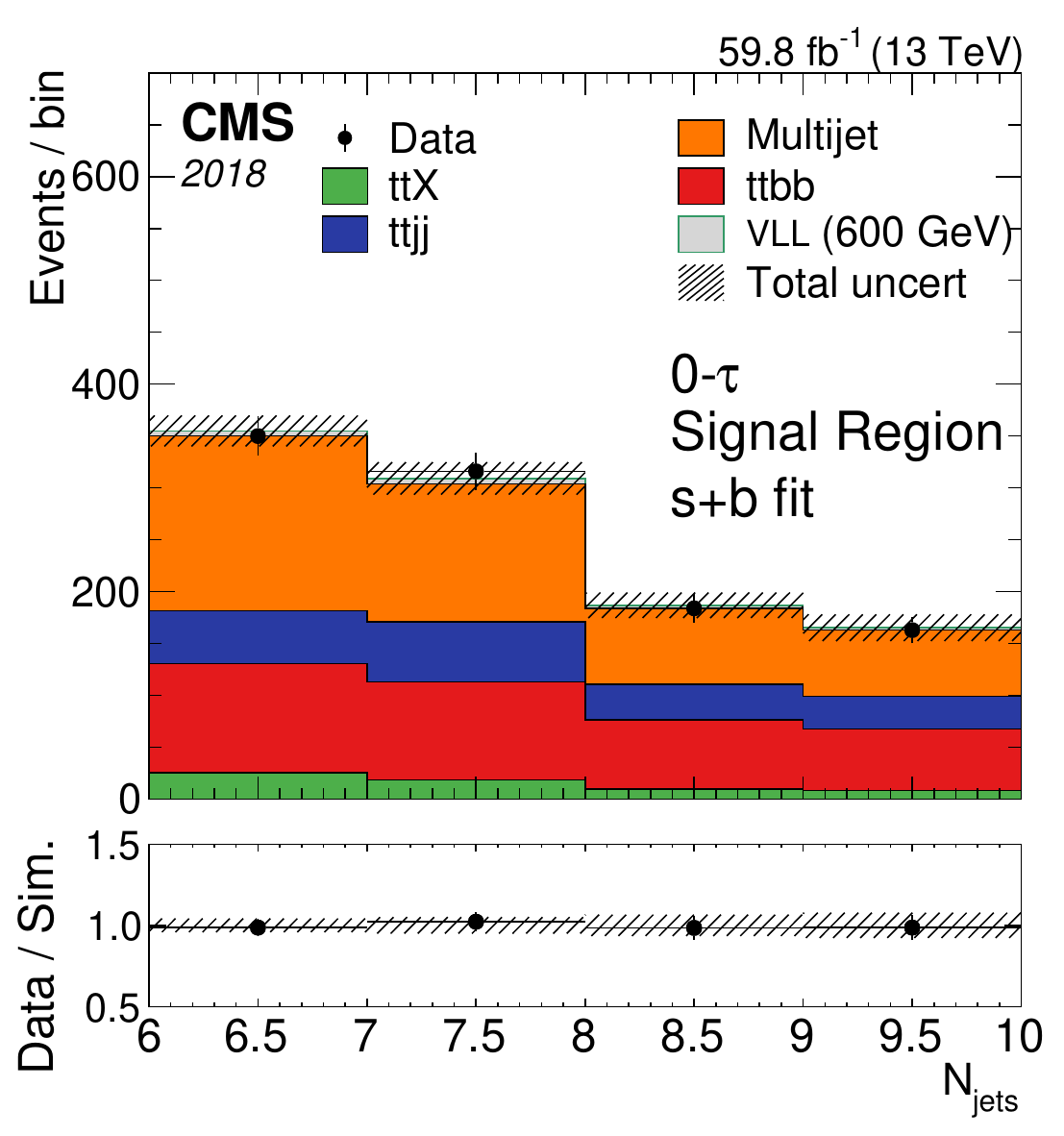}
    \includegraphics[width=0.29\textwidth]{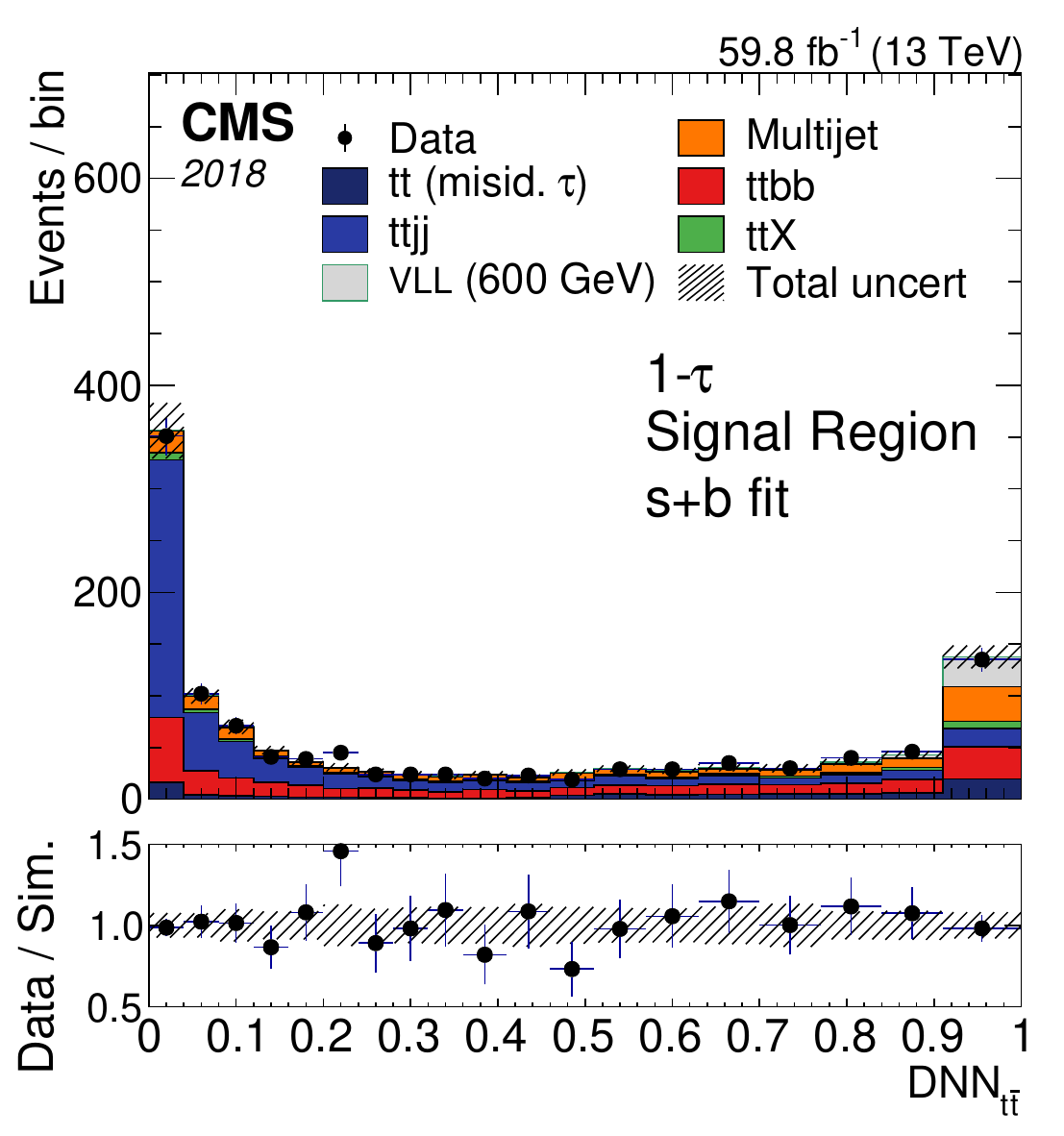}
    \includegraphics[width=0.29\textwidth]{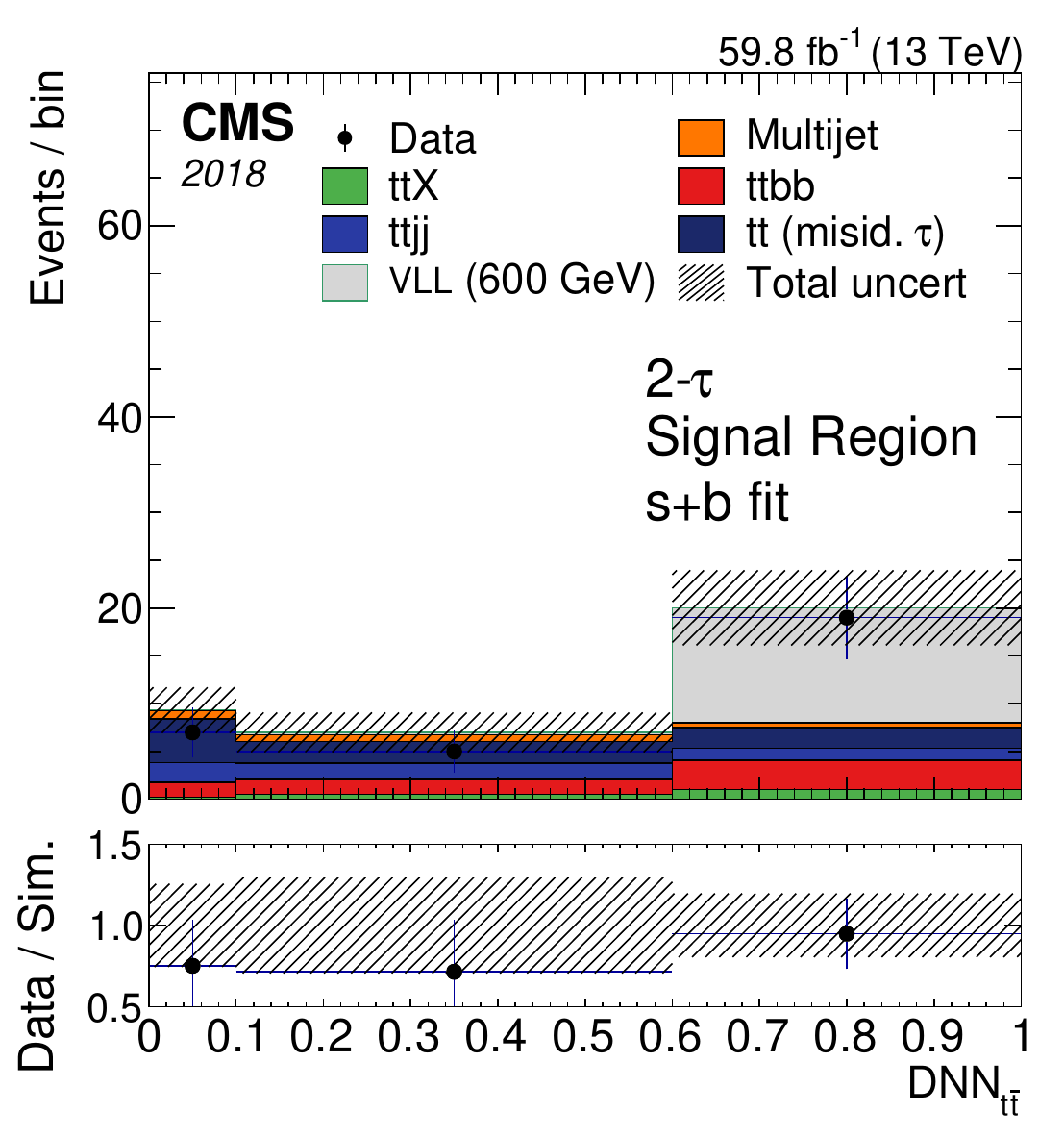}
  
    \caption{Postfit plots for 2017 (top two rows) and 2018 (bottom two rows) showing the fitted distributions in the signal region for the different {\tauh} multiplicity channels (from left to right: 0-{\tauh}, 1-{\tauh}, 2-{\tauh}). The jet multiplicity is fit for the 0-{\tauh} region, whereas the {\DNNTT} score is fit for the 1-{\tauh} and 2-{\tauh} regions. The first and third row show the background-only fits and the second and fourth rows show the fit including the signal.}
    \label{fig:postfit}
\end{figure*}

 The expected and observed limits are shown in Fig.~\ref{fig:limits}. The {\DNNTT} output, used in the 1-{\tauh} and 2-{\tauh} signal regions, is not very sensitive to the signal mass, and therefore the results are highly correlated across all mass points. The higher than expected limits are consistent with the existence of a signal. As a result, none of the mass range tested here (500--1050\GeV) is excluded at the 95\% \CL, and limits are set between 10 and 30\unit{fb}, depending on the \VLL mass hypothesis. At the representative 600\GeV mass point, the signal-plus-background model is preferred over the background-only model at the level of 2.8 standard deviations ($\sigma$). The highest {\DNNTT} bins of each region fit the data better when the signal hypothesis is included, especially for the 2018 2-{\tauh} region, where the background-dominated part of the fit is also substantially improved. Nuisance parameters related to the fake factor estimates and the statistical uncertainties of the simulated events in high {\DNNTT} bins are also pulled to lower likelihood values in the background-only fit, contributing to the significance.

A number of post-unblinding checks were performed to test the robustness of the result, reliability of the model, and consistency of the data with the signal hypothesis. As indicated by the significance in the combined fit, the signal interpretation is consistent across the channels. No excess is observed in the significantly less sensitive 0-{\tauh} regions, whereas the excesses observed in the 1-{\tauh} and 2-{\tauh} regions are consistent with each other. The observed limits for each of the 1-{\tauh} and 2-{\tauh} channels, when fit individually, are above the expected limits. Taking the 600\GeV mass point as an example, the significances for these channels when fit individually range from 0.8 to 2.0$\sigma$ and are all within 1$\sigma$ of their expected significances for a signal strength corresponding to the best combined fit signal strength. 

Several alternative analysis procedures were also tested and found to give similar results to those presented here. These included rejecting {\tauh} candidates with only two reconstructed tracks, applying a reweighting to account for possible mismodelling of the \ttbar \HT spectrum, increasing the size of certain experimental uncertainties, and including an additional uncertainty related to the difference between data and simulation in the {\tauh} FFs. All of these were found to give very similar results to the main result presented here.  

These results have been obtained considering only electroweak production of {\VLL} pairs. However, tests showed that the results hold when including \PZpr production of \VLL pairs as well.

\begin{figure}[htb!]
    \centering
    \includegraphics[width=\cmsFigWidth]{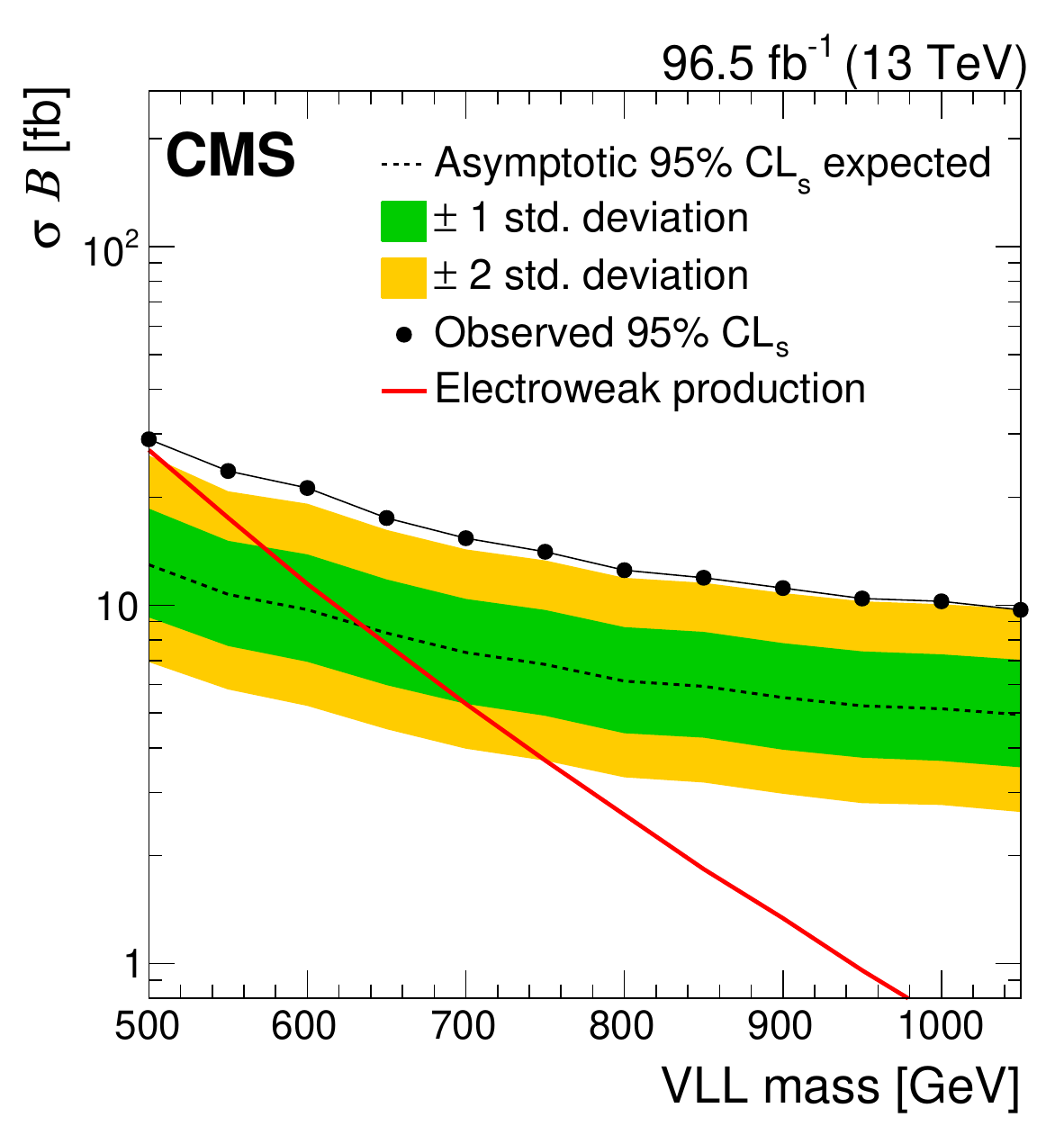}
    \caption{Expected and observed 95\% \CL upper limits on the product of the \VLL pair production cross section and the branching fraction to third generation quarks and leptons, combining the 2017 and 2018 data and all \tauh multiplicity channels. The theoretical prediction in the 4321 model for electroweak production of {\VLL}s is also shown.}
    \label{fig:limits}
\end{figure}

\ifthenelse{\boolean{cms@external}}{}{\clearpage}
\section{Summary}\label{sec:summary}
The first search for vector-like leptons ({\VLL}s) in the context of the 4321 model has been presented, using proton-proton collision data collected with the CMS detector at $\sqrt{s}=13\TeV$, corresponding to an integrated luminosity of 96.5\fbinv. The probed model consists of an extension of the standard model with an $\text{SU}(4) \times \text{SU}(3)' \times \text{SU}(2)_L \times \text{U}(1)'$ gauge sector that can provide a combined explanation for multiple anomalies observed in {\PQb} hadron decays, which point to lepton flavour nonuniversality. In the model, a leptoquark is predicted as the primary source of lepton flavour nonuniversality, while the ultraviolet-completion predicts additional vector-like fermion families. In particular, {\VLL}s are investigated via their couplings to standard model fermions through leptoquark interactions, resulting in third-generation fermion signatures. Final states containing at least three \PQb-tagged jets and varying $\PGt$ lepton multiplicities are considered. To improve the search sensitivity, graph neural networks are trained to discriminate between classes of events using the kinematic relationships between particles in an environment with high jet multiplicity.

An excess of 2.8 standard deviations over the standard model background-only hypothesis was observed in the data at the representative \VLL mass point of 600\GeV. As a result, no \VLL masses are excluded at the 95\% confidence level and limits on the product of the {\VLL} pair production cross section and their branching ratio to third generation fermions are set between 10 and 30\unit{fb}, depending on the \VLL mass hypothesis.

\begin{acknowledgments}
We congratulate our colleagues in the CERN accelerator departments for the excellent performance of the LHC and thank the technical and administrative staffs at CERN and at other CMS institutes for their contributions to the success of the CMS effort. In addition, we gratefully acknowledge the computing centres and personnel of the Worldwide LHC Computing Grid and other centres for delivering so effectively the computing infrastructure essential to our analyses. Finally, we acknowledge the enduring support for the construction and operation of the LHC, the CMS detector, and the supporting computing infrastructure provided by the following funding agencies: BMBWF and FWF (Austria); FNRS and FWO (Belgium); CNPq, CAPES, FAPERJ, FAPERGS, and FAPESP (Brazil); MES and BNSF (Bulgaria); CERN; CAS, MoST, and NSFC (China); MINCIENCIAS (Colombia); MSES and CSF (Croatia); RIF (Cyprus); SENESCYT (Ecuador); MoER, ERC PUT and ERDF (Estonia); Academy of Finland, MEC, and HIP (Finland); CEA and CNRS/IN2P3 (France); BMBF, DFG, and HGF (Germany); GSRI (Greece); NKFIH (Hungary); DAE and DST (India); IPM (Iran); SFI (Ireland); INFN (Italy); MSIP and NRF (Republic of Korea); MES (Latvia); LAS (Lithuania); MOE and UM (Malaysia); BUAP, CINVESTAV, CONACYT, LNS, SEP, and UASLP-FAI (Mexico); MOS (Montenegro); MBIE (New Zealand); PAEC (Pakistan); MES and NSC (Poland); FCT (Portugal); MESTD (Serbia); MCIN/AEI and PCTI (Spain); MOSTR (Sri Lanka); Swiss Funding Agencies (Switzerland); MST (Taipei); MHESI and NSTDA (Thailand); TUBITAK and TENMAK (Turkey); NASU (Ukraine); STFC (United Kingdom); DOE and NSF (USA).

\hyphenation{Rachada-pisek} Individuals have received support from the Marie-Curie programme and the European Research Council and Horizon 2020 Grant, contract Nos.\ 675440, 724704, 752730, 758316, 765710, 824093, 884104, and COST Action CA16108 (European Union); the Leventis Foundation; the Alfred P.\ Sloan Foundation; the Alexander von Humboldt Foundation; the Belgian Federal Science Policy Office; the Fonds pour la Formation \`a la Recherche dans l'Industrie et dans l'Agriculture (FRIA-Belgium); the Agentschap voor Innovatie door Wetenschap en Technologie (IWT-Belgium); the F.R.S.-FNRS and FWO (Belgium) under the ``Excellence of Science -- EOS" -- be.h project n.\ 30820817; the Beijing Municipal Science \& Technology Commission, No. Z191100007219010; the Ministry of Education, Youth and Sports (MEYS) of the Czech Republic; the Hellenic Foundation for Research and Innovation (HFRI), Project Number 2288 (Greece); the Deutsche Forschungsgemeinschaft (DFG), under Germany's Excellence Strategy -- EXC 2121 ``Quantum Universe" -- 390833306, and under project number 400140256 - GRK2497; the Hungarian Academy of Sciences, the New National Excellence Program - \'UNKP, the NKFIH research grants K 124845, K 124850, K 128713, K 128786, K 129058, K 131991, K 133046, K 138136, K 143460, K 143477, 2020-2.2.1-ED-2021-00181, and TKP2021-NKTA-64 (Hungary); the Council of Science and Industrial Research, India; the Latvian Council of Science; the Ministry of Education and Science, project no. 2022/WK/14, and the National Science Center, contracts Opus 2021/41/B/ST2/01369 and 2021/43/B/ST2/01552 (Poland); the Funda\c{c}\~ao para a Ci\^encia e a Tecnologia, grant CEECIND/01334/2018 (Portugal); the National Priorities Research Program by Qatar National Research Fund; MCIN/AEI/10.13039/501100011033, ERDF ``a way of making Europe", and the Programa Estatal de Fomento de la Investigaci{\'o}n Cient{\'i}fica y T{\'e}cnica de Excelencia Mar\'{\i}a de Maeztu, grant MDM-2017-0765 and Programa Severo Ochoa del Principado de Asturias (Spain); the Chulalongkorn Academic into Its 2nd Century Project Advancement Project, and the National Science, Research and Innovation Fund via the Program Management Unit for Human Resources \& Institutional Development, Research and Innovation, grant B05F650021 (Thailand); the Kavli Foundation; the Nvidia Corporation; the SuperMicro Corporation; the Welch Foundation, contract C-1845; and the Weston Havens Foundation (USA).    
\end{acknowledgments}
\bibliography{auto_generated} 
\cleardoublepage \appendix\section{The CMS Collaboration \label{app:collab}}\begin{sloppypar}\hyphenpenalty=5000\widowpenalty=500\clubpenalty=5000
\cmsinstitute{Yerevan Physics Institute, Yerevan, Armenia}
{\tolerance=6000
A.~Tumasyan\cmsAuthorMark{1}\cmsorcid{0009-0000-0684-6742}
\par}
\cmsinstitute{Institut f\"{u}r Hochenergiephysik, Vienna, Austria}
{\tolerance=6000
W.~Adam\cmsorcid{0000-0001-9099-4341}, J.W.~Andrejkovic, T.~Bergauer\cmsorcid{0000-0002-5786-0293}, S.~Chatterjee\cmsorcid{0000-0003-2660-0349}, K.~Damanakis\cmsorcid{0000-0001-5389-2872}, M.~Dragicevic\cmsorcid{0000-0003-1967-6783}, A.~Escalante~Del~Valle\cmsorcid{0000-0002-9702-6359}, P.S.~Hussain\cmsorcid{0000-0002-4825-5278}, M.~Jeitler\cmsAuthorMark{2}\cmsorcid{0000-0002-5141-9560}, N.~Krammer\cmsorcid{0000-0002-0548-0985}, L.~Lechner\cmsorcid{0000-0002-3065-1141}, D.~Liko\cmsorcid{0000-0002-3380-473X}, I.~Mikulec\cmsorcid{0000-0003-0385-2746}, P.~Paulitsch, F.M.~Pitters, J.~Schieck\cmsAuthorMark{2}\cmsorcid{0000-0002-1058-8093}, R.~Sch\"{o}fbeck\cmsorcid{0000-0002-2332-8784}, D.~Schwarz\cmsorcid{0000-0002-3821-7331}, S.~Templ\cmsorcid{0000-0003-3137-5692}, W.~Waltenberger\cmsorcid{0000-0002-6215-7228}, C.-E.~Wulz\cmsAuthorMark{2}\cmsorcid{0000-0001-9226-5812}
% ...
\par}
\vskip\cmsinstskip
\dag:~Deceased\\
$^{1}$Also at Yerevan State University, Yerevan, Armenia\\
$^{2}$Also at TU Wien, Vienna, Austria\\
% ...
\end{sloppypar}
\end{document}